\documentclass{article}
\PassOptionsToPackage{numbers, sort, compress}{natbib}

\usepackage[preprint]{neurips_2024}

\usepackage[utf8]{inputenc} 
\usepackage[T1]{fontenc}    
\usepackage{hyperref}       
\usepackage{url}            
\usepackage{booktabs}       
\usepackage{amsmath}        
\usepackage{amsfonts}       
\usepackage{nicefrac}       
\usepackage{microtype}      
\usepackage[table]{xcolor}  
\usepackage{xcolor}         
\usepackage{lipsum}
\usepackage{graphicx}
\usepackage{multirow}
\usepackage{graphics}
\usepackage{wrapfig}
\usepackage{caption}
\usepackage{algorithm}
\usepackage{algpseudocode}
\usepackage{enumitem}
\usepackage{array}

\newcommand{\pxxx}{\makebox[0pt][r]{$^{***}$}}
\newcommand{\pxx}{\makebox[0pt][r]{$^{**}$}}
\newcommand{\px}{\makebox[0pt][r]{$^{*}$}}
\newcommand{\ci}[1]{{\tiny $\pm${#1}}}

\newcommand{\R}{\mathbb R}
\newcommand{\N}{\mathbb N}
\newcommand{\U}{\mathcal U}

\newcommand{\nN}{\mathcal N}
\newcommand{\dd}{\mathrm d}
\newcommand{\what}{\widehat}

\newcommand{\dpkl}[2]{\left (#1\middle\| #2\right)}
\newcommand{\dkl}[2]{D_{\text{KL}}\dpkl{#1}{#2}}
\newcommand{\dikl}[2]{D_{\text{IKL}}\dpkl{#1}{#2}}

\newcommand{\norm}[1]{\left\|{#1}\right\|}
\DeclareMathOperator*{\argmin}{arg\,min}
\newcommand{\p}[1]{\left({#1}\right)}
\newcommand{\floor}[1]{\left\lfloor{#1}\right\rfloor}
\newcommand{\cb}[1]{\left \{{#1}\right\}}
\newcommand{\bb}[1]{\left [{#1}\right]}

\title{Autoregressive Diffusion Transformer for Text-to-Speech Synthesis}

\author{%
  Zhijun Liu$^{1,\dagger}$\quad~Shuai Wang$^{2,1,\ddagger}$\quad~Sho~Inoue$^1$\quad~Qibing~Bai$^1$\quad~Haizhou~Li$^{1,2}$\\
  $^1$School of Data Science, $^2$Shenzhen Research Institute of Big Data\\
  The Chinese University of Hong Kong, Shenzhen, Guangdong, P.R. China\\
}

\begin{document}
\maketitle
\begin{abstract}
Audio language models have recently emerged as a promising approach for various audio generation tasks, relying on audio tokenizers to encode waveforms into sequences of discrete symbols. Audio tokenization often poses a necessary compromise between code bitrate and reconstruction accuracy.
When dealing with low-bitrate audio codes, language models are constrained to process only a subset of the information embedded in the audio, which in turn restricts their generative capabilities.
To circumvent these issues, we propose encoding audio as vector sequences in continuous space $\R^d$ and autoregressively generating these sequences using a decoder-only diffusion transformer (ARDiT). Our findings indicate that ARDiT excels in zero-shot text-to-speech and exhibits performance that compares to or even surpasses that of state-of-the-art models. High-bitrate continuous speech representation enables almost flawless reconstruction, allowing our model to achieve nearly perfect speech editing.
Our experiments reveal that employing Integral Kullback-Leibler (IKL) divergence for distillation at each autoregressive step significantly boosts the perceived quality of the samples. Simultaneously, it condenses the iterative sampling process of the diffusion model into a single step.
Furthermore, ARDiT can be trained to predict several continuous vectors in one step, significantly reducing latency during sampling. Impressively, one of our models can generate $170$ ms of $24$ kHz speech per evaluation step with minimal degradation in performance.
Audio samples are available at \href{https://ardit-tts.github.io/}{\nolinkurl{ardit-tts.github.io}}.
\end{abstract}

\renewcommand{\thefootnote}{$\dagger$}
\footnotetext[\value{footnote}]{Email: \href{mailto:zhijunliu1@link.cuhk.edu.cn}{\nolinkurl{zhijunliu1@link.cuhk.edu.cn}}}
\renewcommand{\thefootnote}{$\ddagger$}
\footnotetext[\value{footnote}]{Corresponding to: Shuai Wang (\href{mailto:wangshuai@cuhk.edu.cn}{\nolinkurl{wangshuai@cuhk.edu.cn}})}
\renewcommand{\thefootnote}{\arabic{footnote}}

\section{Introduction}

Autoregressive modeling of discrete audio tokens has recently achieved significant success across various audio generation tasks \cite{AudioLMSurvey,AudioLM,MusicLM,VALLE,AudioGen,VIOLA,MusicGen,AudioPaLM,SpeechX,LauraGPT,UniAudio,GSLM,TWIST,UnitY,SPEARTTS}. These models, often referred to as audio language models, compress audio waveforms into \textit{discrete tokens}, predict them autoregressively, and then decode them back into waveforms. By discretizing audio signals into discrete tokens \cite{SoundStream,EnCodec,AudioDec,HifiCodec,DAC,SpeechTokenizer,FunCodec}, a unified representation of audio and text is achieved, enabling  seamless joint processing with language models.

Despite its success, discrete audio tokenization faces challenges. The theory of lossy compression \cite{RDPTradeoff,PDTradeoff} suggests \textit{a trade-off between the bitrate and reconstruction quality}. Current state-of-the-art neural audio codecs typically require a minimum of $1.5$ kbps for high-fidelity reconstruction of $16$ kHz audio~\cite{AudioLMSurvey}. Using a codebook of size $1024$, one second of audio would necessitate $1500 / \log_2(1024) = 150$ tokens, leading to long sequences that complicate audio language modeling. Different strategies have been proposed to mitigate this, each with its own limitations:
\textbf{(i)}~A prevalent approach assumes conditional independence among tokens \cite{VALLE,AudioGen,MusicGen}, concurrently generating multiple tokens to reduce autoregressive sampling steps. However, the effectiveness of this assumption depends on the quantization method employed. For example, Delayed Pattern~\cite{MusicGen,VoiceCraft} works well with residual vector quantization (RVQ) \cite{EnCodec}, but it might not work well with other techniques such as finite scalar quantization (FSQ) \cite{FSQ}. An inappropriate independence assumption can negatively impact generation performance \cite{MusicGen}.
\textbf{(ii)}~Another approach involves limiting the bitrate and shortening the token sequence length by encoding only a fraction of the total information in audios \cite{AudioLM,MusicLM,AudioPaLM,GSLM,DiscreteResynthesis,TWIST,UnitY,SPEARTTS}. This requires sophisticated disentangled representation learning to achieve a compressed, task-specific representation for the audio language model. And it limits the model to generating only partial audio information, hindering its performance on tasks that require high output bitrate. The information gap must be filled by a cascade of generative models, complicating the audio generation system.

Besides the trade-off between bitrate and reconstruction quality, audio tokenization also faces challenges with \textit{gradient-based optimization in discrete distributions}. While VAEs~\cite{VAE} and VAE-GANs~\cite{VAEGAN} with continuous latent variables can be trained using standard gradient optimizers via the reparameterization trick~\cite{VAE}, this is not the case for VQ-GAN models~\cite{VQGAN}, which form the basis of modern neural audio codecs. Effectively training VQ-GANs necessitates techniques like auxiliary losses and codebook re-initialization \cite{VQVAE,Wav2Vec2,StraightThrough,GumbelSoftmax,VQWav2Vec}.

The complexities of audio tokenization can be avoided by representing speech as sequences of vectors in $\R^d$, termed as \textit{continuous tokens}~\cite{GSLMCont}.
For a continuous token sequence $[x_1, \cdots, x_N]$, several methods have been explored to model its conditional density $p(x_n | x_{<n})$:
\textbf{(i)}~One approach is to apply flows based on finite compositions~\cite{NormalizingFlow, Glow,WaveTacotron,Flowtron,WaveFlow} for $p_\theta(x_n|x_{<n})$. This model constrains the network architecture to ensure efficient computation of the Jacobian determinant and thereby guarantee invertibility.
\textbf{(ii)}~An alternative is to represent $p_\theta(x_n|x_{<n})$ using a Mixture Density Network (MDN)~\cite{MDN} that predicts parameters for continuous mixture distributions \cite{MDNTTSDu,MDNTTSWang,CLaMTTS,PixelCNN++,ParallelWaveNet,GIVT,Tacotron}, such as a mixture of Gaussians (MoG). However, due to the limited expressive power of mixture densities, $p(x_n | x_{<n})$ needs to be simple enough to be accurately approximated.
\textbf{(iii)}~Efforts to model $p_\theta(x_{n} | x_{<n})$ with generative adversarial networks (GANs)~\cite{GAN} have been made~\cite{ChunkWaveGAN}, but they often suffer from training instability and mode dropping issues.
\textbf{(iv)}~Diffusion probabilistic models (DPMs)~\cite{DPMZero,DDPM,ScoreSDE} are proficient at modeling continuous densities. Their integration with autoregressive models could yield impressive results~\cite{SSDLM,SSDLM2,GroupwiseDiffusion,HierMusic,Jen1,VASA1,DiffAR,LAVIE,VDT}. However, DPMs require iterative processes for generating high-quality samples. Combining diffusion sampling with autoregressive sequence sampling can result in a slow, high-latency sampling process.

Recent discoveries in the distillation of diffusion models~\cite{DiffInstruct,ScoreGAN,VSD,DMD,SiD} have changed the situation. A family of diffusion distillation methods demonstrates that we can effectively transform diffusion models into single-step implicit generative models while preserving or even improving their generative modeling performance. SSD-LM~\cite{SSDLM, SSDLM2} developed a method of integrating autoregressive sequence modeling with diffusion models utilizing a decoder-only transformer. Compared to SSD-LM \cite{SSDLM}, SSD-2 \cite{SSDLM2} carefully designed the attention mask that enhances the training efficiency of autoregressive diffusion transformers (ARDiTs). However, it still suffers from slow speed and high computational cost during inference. In our study, we propose to apply Distribution Matching Distillation (DMD) \cite{DiffInstruct,DMD} to distill an autoregressive diffusion transformer for audio generation. To verify the performance of this integrated approach, we apply it to zero-shot text-to-speech synthesis and speech editing tasks.

Our contributions can be summarized as follows:
\begin{itemize}
    \item We introduce ARDiT for audio generation, a decoder-only diffusion transformer model that eliminates the need for discrete tokenization of audio signals.
    \item Leveraging fill-in-the-middle (FIM) \cite{FIM} training, ARDiT excels in zero-shot text-to-speech synthesis and speech editing, showcasing near-perfect speech editing capabilities on the LibriTTS dataset.
    \item We distill ARDiT text-to-speech (TTS) models with DMD. After distillation, the student models demonstrate enhanced perceptual naturalness compared to the teacher models, while requiring only one network evaluation to generate one or more continuous tokens. One of our distilled models achieves speech generation speeds of 170ms per network evaluation, significantly reducing the inference latency.
    \item Furthermore, we present a novel method for controlling the total duration of generated speech in ARDiT TTS by manipulating the rotation angles of Rotary Position Embeddings (RoPE) \cite{Roformer}.
\end{itemize}

\section{Related Works}

\textbf{Autoregressive Diffusion Models}: 
Autoregressive diffusion models decompose high-dimensional data into lower-dimensional segments. These segments are then sequentially generated using diffusion models. These models have been successfully applied across various domains, including image~\cite{GroupwiseDiffusion}, video~\cite{VideoDiffusionModels, RollingDiffusionModels, VASA1}, music~\cite{HierMusic, Jen1}, speech~\cite{DiffAR}, and text~\cite{SSDLM,SSDLM2}. For instance, in the music sector, the studies by \cite{HierMusic} and \cite{Jen1} focus on long-term generation and efficiency enhancement, respectively. In the speech domain, \cite{DiffAR} develops a model for generating a speech waveform from phonemes, durations, and pitch contours.

\textbf{Zero-shot TTS}:
Unlike traditional speech synthesis techniques that require hours of high-quality transcribed data from the target speaker, zero-shot TTS aims to customize a new speaker's voice with just a few seconds of a voice prompt. Previously, zero-shot TTS was typically achieved using a pretrained speaker encoder~\cite{TacotronSpeaker}, which could only retain partial information about the speaker's timbre and failed to capture other aspects such as style. Currently, thanks to new model frameworks and scaled datasets, zero-shot TTS has made significant progress. Solutions can be divided into two main categories: the first is based on autoregressive codec language models, such as Vall-E~\cite{VALLE}, Vall-E~X~\cite{VALLEX}, and SpearTTS~\cite{SPEARTTS}. The other type relies on non-autoregressive generative models, such as VQ-GAN based MegaTTS~\cite{MegaTTS} and MegaTTS 2~\cite{MegaTTS2}  , diffusion-based UniCATS~\cite{UniCATS}, NaturalSpeech2~\cite{NaturalSpeech2}, NaturalSpeech3~\cite{NaturalSpeech3}, and VoiceBox~\cite{VoiceBox}.

\textbf{Text-Based Speech Editing}: Text-based speech editing adjusts segments of an utterance to fit a target transcript while preserving the unaltered parts. Early methods, such as those by~\cite{Voco}, employed a TTS and voice conversion model for text-guided speech changes, but often resulted in unnatural speech due to prosody and boundary discrepancies. More recent models~\cite{EditSpeech,CampNet,A3T,SpeechPainter,FluentSpeech,UniCATS,DiffVoice,EdiTTS,RetrieverTTS,E3TTS,VoiceCraft,VoiceBox} have sought to enhance naturalness by conditioning the generation on the surrounding speech context.

\textbf{Diffusion Distillation Methods}: 
In the context of diffusion models, distillation is typically employed to reduce the number of sampling steps during inference \cite{BlogDistillation}. One family of distillation methods~\cite{DirectDD, ProgressiveDistillation, ConsistencyModel, TRACT, BOOT, DSNO, TCD} seeks to preserve the bijection from noise to data defined by probability flow ODEs \cite{DDIM, ScoreSDE} in diffusion models. Another family of distillation methods \cite{SiD, DiffInstruct, DMD, ScoreGAN, VSD} distills diffusion models into one-step generators by minimizing the divergence between the distributions of generated samples and data. These diffusion distillation techniques have been investigated in diffusion-based TTS systems, resulting in several efficient TTS models \cite{ProDiff, ReflowTTS, CoMoSpeech, DiffGANTTS}.

\section{Method}

\subsection{Background: Flow Matching and Distribution Matching Distillation}

\label{sec:flow}

Suppose $X, Z$ are independent $\R^d$-valued random variables with data density $p(x)$ and Gaussian density $p(z) = \nN(0, I_d)$. Let $\alpha_t = (1 - t)$ and $\sigma_t = t$ for $t \in [0, 1]$. Let $X_t = \alpha_t X + \sigma_t Z$. Define the velocity field $v(x_t, t): \R^{d} \times [0, 1] \to \R^d$ as:
\begin{equation}
v(x_t, t) := \argmin_{v}E\norm{v(X_t, t) - (Z-X)}_2^2 = E[Z - X \mid X_t = x_t].
\end{equation}
According to \cite{Reflow,FlowMatching,StochasticInterpolants}, we can sample $p(x)$ by solving the following ODE in reverse:
\begin{equation}
\dd Y_t = v(Y_t, t) \dd t, \quad Y_1 \sim \nN(0, I_d), \quad t \in [0, 1].
\label{equ:flowode}
\end{equation}
Therefore we can obtain a deep generative model with sample density $p_\theta(x) \approx p(x)$ by estimating $v(x_t, t)$ with $v_\theta(x_t, t)$ through minimizing $E_{t \sim \U[0, 1]}\norm{v_\theta(X_t, t) - (Z - X)}_2^2$.
This generative model is referred to by various names in the literature \cite{Reflow,FlowMatching,StochasticInterpolants,SiT}. In the following discussion, we will refer to it as "Flow Matching" \cite{FlowMatching}.

Suppose $X_t \sim p_t(x_t)$. We can show that the score function $s(x_t, t) = \nabla_{x_t} \log p_t(x_t)$ can be extracted from the velocity field $v(x_t, t)$ (See Appendix \ref{appendix:flow_score}). 
\begin{equation}
v(x_t, t) = -\sigma_t s(x_t, t) - \frac{x_t + \sigma_t^2 s(x_t, t)}{\alpha_t} = \frac{-1}{1 - t}x_t + \frac{-t}{1 - t} s(x_t, t).
\label{equ:v_s_equivalent}
\end{equation}
This indicates that, akin to Diffusion Probabilistic Models (DPMs) \cite{DDPM,ScoreSDE}, Flow Matching models also estimate the score function.

Given a Flow Matching model $v_\theta(x_t, t)$ trained on $p(x)$. With the ODE in equation \ref{equ:flowode}, it establishes a mapping $f_\theta(w): \R^d \to \R^d$ that transforms Gaussian noises to data samples. Evaluating $f_\theta$ is slow as it involves solving an ODE. DMD can distill $v_\theta$ into single step generator $g_\xi: \R^d \to \R^d$, that maps random noise $W \sim \nN(0, I_d)$ to $\what X = g_\xi(W)$ with density $p_\xi(x)$. Define $\what X_t := \alpha_t \what X + \sigma_t \what Z$ where $\what Z$ is an independent Gaussian random variable. Suppose $p(x_t, t)$ is the density of $X_t$ and $p_\xi(x_t, t)$ is the density of $\what X_t$. Their Integral Kullback–Leibler (IKL) divergence \cite{DiffInstruct} is defined as:
\begin{equation}
D_\xi := \dikl{p_\xi(x_t, t)}{p(x_t, t)} := E_{t \sim \U[0, 1]}\bb{w_t \dkl{p_{\xi}(x_t, t)}{p(x_t, t)}},
\end{equation}
where $w_t \ge 0$ is the weighting factor for time $t$. Suppose $s(x_t, t) = \nabla_{x_t} \log p(x_t, t)$ and $s_\xi(x_t, t) := \nabla_{x_t} \log p_\xi(x_t, t)$. Then according to \cite{DiffInstruct, DMD}:
\begin{equation}
\nabla_\xi D_\xi = E_{t \sim \U[0, 1]}\bb{w_t \alpha_t \p{s_\xi(\what X_t, t) - s(\what X_t, t)} \frac{\partial g_\xi(W)}{\partial\xi}}.
\end{equation}
$s_\xi(x_t, t)$ and $s(x_t, t)$ are unknown, but we can approximate $s_\xi(x_, t) - s(x_t, t)$ with Flow Matching models $v_\eta(x_t, t)$ and $v_\theta(x_t, t)$. Where $v_\eta$ is trained on samples of $g_\xi$ by minimizing:
\begin{equation}
\mathcal L_\eta := E_{t \sim \U[0, 1]}\norm{v_\eta(\what X_t, t) - (\what Z - \what X)}_2^2.
\label{equ:fm_fake}
\end{equation}
According to equation \ref{equ:v_s_equivalent}, assuming $v_\eta$ and $v_\theta$ are well-trained, we have:
\begin{equation}
v_\eta(x_t, t) - v_\theta(x_t, t) \approx \frac{-t}{1-t} \cdot \p{s_\xi(x_t, t) - s(x_t, t)}.
\end{equation}
DMD training freezes $v_\theta$ and alternatively updates $g_\xi$ and $v_\eta$ with $\mathcal L_\xi$ and $\mathcal L_\eta$. The generator $g_\xi$ is trained on minimizing the following loss function:
\begin{equation}
\mathcal L_\xi := \underbrace{E_{t \sim \U[0, 1]}\norm{\what X +\operatorname{sg}\p{v_\theta(\what X_t, t) - v_\eta(\what X_t, t) - \what X}}_2^2}_{\mathcal L_{\text{IKL}}} + \beta_{\text{reg}} \cdot \underbrace{E \norm{g_\xi(W) - f_\theta(W)}_2^2}_{\mathcal L_{\text{reg}}}.
\label{equ:dmd_general}
\end{equation}
Here $\beta_{\text{reg}} > 0$ is the weight of the L2 regression loss. And $\operatorname{sg}$ is the stop gradient operator. For simplicity, we set $w_t \alpha_t = 2t / (1 - t)$. In this case $\nabla_\xi \mathcal L_{\text{IKL}} \approx \nabla_\xi D_\xi$. Impact of the weighting factor $w_t$ is left for future study.

\subsection{Contextual Mel Spectrogram Autoencoder}

\label{sec:melctx}

\begin{figure}[ht]
    \centering
    \includegraphics[width=1.0\textwidth,trim=.6cm .2cm .6cm .2cm,clip]{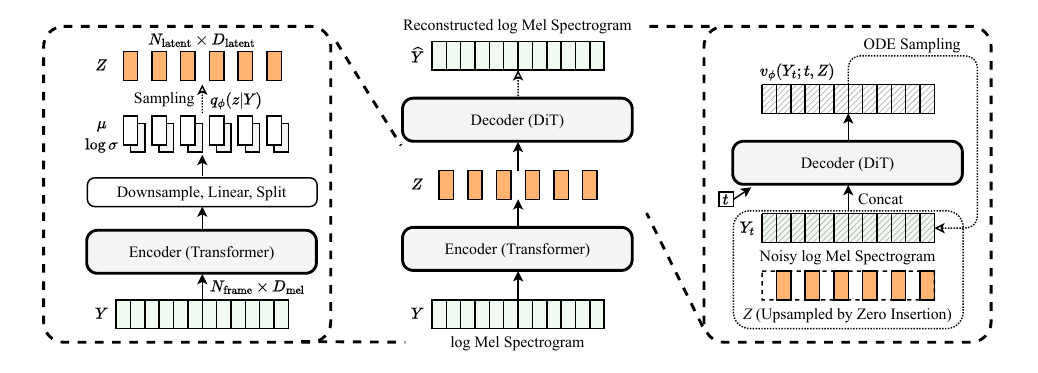}
    \caption{The structure of the Mel spectrogram autoencoder. The left-hand side illustrates the encoder, which outputs the mean and variance of the encoder distribution $q_\phi(z|Y)$. The right-hand side illustrates the Flow Matching decoder. The encoder and decoder are jointly trained to minimize the reconstruction error and the bitrate.}
    \label{fig:ardit_ae}
\end{figure}

We compress log Mel spectrograms into a sequence of continuous tokens with an autoencoder to reduce the sequence length. Given random Mel spectrogram $Y$ on $\R^{N_{\text{frame}} \times D_{\text{mel}}}$ where $N_{\text{frame}}$ is the number of frames, and $D_{\text{mel}}$ is the number of Mel filters, we encode $Y$ into a sequence of continuous tokens $Z$ on $\R^{N_{\text{latent}} \times D_{\text{latent}}}$ where $N_{\text{latent}} = \floor{N_{\text{frame}} / 4}$ and $D_{\text{latent}} = 16$. The encoder is a transformer~\cite{Transformer} taking input $Y$ and outputs $\mu, \log \sigma \in \R^{N_{\text{latent}} \times D_{\text{latent}}}$. The encoder defines the conditional density of $Z$ given $Y$ as $q_{\phi}(z|y) = \prod_{n, d}\nN(z_{n, d}; \mu_{n, d}, \sigma_{n, d}^2)$. The decoder is a conditional Flow Matching model $v_\psi(y_t; t, z)$ based on DiT~\cite{DiT} that recovers $Y$ given $Z$. Define the latent prior density $p(z) := \prod_{n, d} \nN(z_{n, d}; 0, 1)$ on $\R^{N_{\text{latent}} \times D_{\text{latent}}}$. The encoder and decoder are jointly optimized by minimizing:
\begin{equation}
\mathcal L(\phi, \psi) := \beta_{\text{MI}} \cdot E\bb{\dkl{q_\phi(z|Y)}{p(z)}} + E_{W \sim \nN(0, I)}\norm{v_\psi\p{\p{1-t}Y + t W; t, Z} - \p{W - Y}}_2^2.
\label{equ:mel_vae}
\end{equation}
Note that the first term $E\bb{\dkl{q_\phi(z|Y)}{p(z)}}$ is a variational upper bound \cite{MIBounds} of mutual information $I(Y; Z)$. So the weight $\beta_{\text{MI}}>0$ is controlling the trade-off between the coding rate and reconstruction accuracy. In our experiments, the Mel spectrogram encoder emits 23.5 tokens per second, and its theoretical bitrate is 1.7 kbps. For more details, please refer to Appendix \ref{appendix:bitrate}.

To enable conditional decoding when the Mel spectrogram target is partially known, the decoder is fine-tuned on Mel spectrogram masked reconstruction. For more details, please refer to Appendix \ref{appendix:mel_inpaint}. During the inference stage of speech editing and zero-shot TTS, we provide the decoder with the known Mel spectrogram frames.

\subsection{Autoregressive Diffusion Transformers for Text-to-Speech Synthesis}

\label{sec:ardit}

In this part, we describe Autoregressive Diffusion Transformers (ARDiTs), and explain how they can be utilized for text-to-speech synthesis. Suppose random Mel spectrogram $Y$ is encoded into continuous token sequence $Z = [Z_0; \cdots; Z_{N_{\text{latent}} - 1}]$ on $\R^{N_{\text{latent}} \times D_{\text{latent}}}$. Suppose $C = [C_0, \cdots, C_{N_{\text{phone}} -1 }]$ on $\Sigma^{N_{\text{phone}}}$ is the phonetic transcript of $Y$. Where $\Sigma$ is the set of all phonemes.

An ARDiT is semi-autoregressive, it samples from conditional density $p_\theta(z^{i:i+B} | c, z^{<i})$ with Flow Matching through estimating the conditional velocity field $v_\theta\p{z_t^{i:i+B}; t, c, z^{<i}}$. Here $B \in \N_+$ is the block size, and $i \in \N_+$ is the index of the first token in block $z^{i: i + B} = \bb{z^i; \cdots; z^{i + B - 1}}$. Suppose $W$ is an independent $\R^{N_{\text{latent}} \times D_{\text{latent}}}$-valued random variable with density $p(w) = \prod_{n, d} \nN(w_{n, d}; 0, 1)$. Let $Z_t = (1-t) Z + t W$. The training loss of ARDiT would be:
\begin{equation}
\mathcal L(\theta) := E_{i , t \sim \U[0, 1]}\norm{v_\theta\p{Z_t^{i:i+B}; t, C, Z^{<i}} - \p{W^{i:i+B} - Z^{i:i+B}}}_2^2.
\end{equation}
A naive implementation of ARDiT would be to feed $(C, Z^{<i}, Z_t^{i: i + B})$ into a transformer with no attention mask, and then combine the last $B$ vector in the output to obtain $v_\theta(Z_t^{i:i+B}; t, C, Z^{<i})$. This implementation is inefficient compared to language models (LMs) on discrete tokens in both training and sampling. During training, it does not support teacher-forcing. $\nabla_\theta \mathcal L(\theta)$ from each batch only depends on a small segment of length $B$ in the model's output. During inference, it does not support KV-cache, causing unnecessary recomputation.

As proposed in SSD-2~\cite{SSDLM2}, a special design of the input sequence and attention mask of the transformer can resolve these performance issues. First, let's split $Z$ into blocks of size $B$. Define block index of $i$th token as $\#_i := \floor{\p{i + S}/{B}}$ where $S \in \cb{0, \cdots, B - 1}$ is an integer constant denoting the block shift. 
Suppose we have $M$ blocks. For each block $m$, $Z^{b_m:e_m}$ represents the tokens within that block, where $b_m$ and $e_m$ denote the beginning and the end indices of the block, respectively.
For each block $m$, pick time $t_m \in [0, 1]$. Let $\mathbf t := [t_0, \cdots, t_{M-1}]$. Define $Z_{\mathbf t}$, where $Z_{\mathbf t}^{b_m:e_m} := (1 - t_m) Z^{b_m :e_m} + t_m W^{b_m :e_m}$ in each block $m$.

\begin{figure}[h]
 \centering
   \includegraphics[width=0.9\textwidth,trim=.3cm .2cm 2.3cm 0cm,clip]{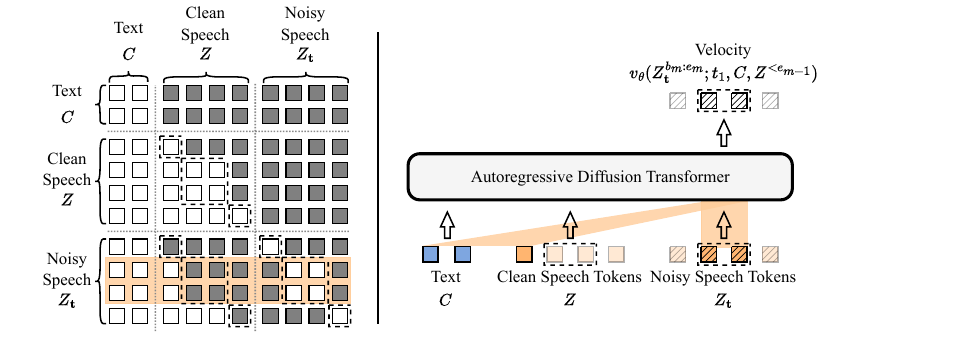}
 \caption{Illustration of the ARDiT training scheme with $B=2$ and $S=1$. The attention mask is depicted on the left, while the right displays the input and output of the ARDiT model during training. The input sequence is divided into three blocks, where noisy speech tokens in each block attend to text tokens, prior clean tokens, and noisy tokens within the same block.}
 \label{fig:attn_train}
\end{figure}

The \textbf{ARDiT training scheme}, including the sequence layout and attention mask during training, is illustrated in figure~\ref{fig:attn_train}. We concat and feed $(C, Z, Z_{\mathbf t})$ into the transformer. The attention mask is defined by the following rules: Tokens in $C$ can attend tokens in $C$; tokens in $Z$ can attend $C$ and tokens in $Z$ with lower or equal block indices; tokens in $Z_{\mathbf t}$ can attend tokens in $Z_{\mathbf t}$ of the same block index and tokens in $Z$ with lower block indices. This training scheme \cite{SSDLM2} is essential for training efficiency. It allows us to evaluate the velocity field $v_\theta(Z_\mathbf t^{b_m:e_m}; t_{m}, C, Z^{<b_m})$ on all the blocks in $Z_{\mathbf{t}}$ with a single neural network evaluation.

\begin{figure}[h]
 \centering
 \includegraphics[width=0.9\textwidth,trim=.6cm .2cm 1.6cm 0cm,clip]{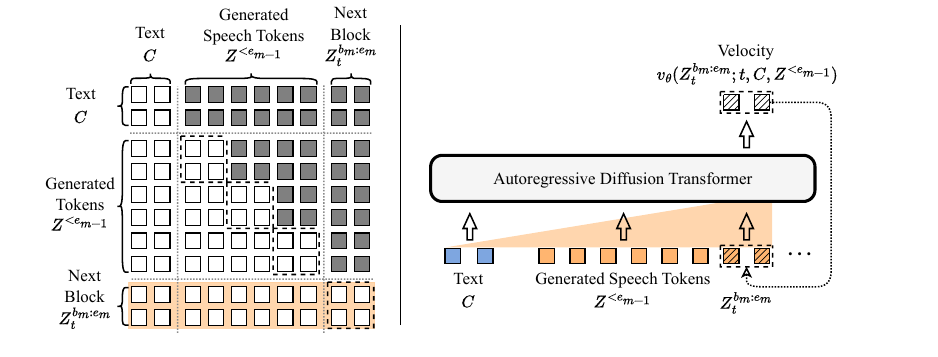}
 \caption{Depiction of the ARDiT inference scheme with $B = 2$. The attention mask is shown on the left, while the right presents the input and output of the ARDiT model during inference. The blocks are generated autoregressively, with the model shown in the process of generating the 4th block of tokens, after already producing 3 blocks of speech tokens. The last block is iteratively sampled by solving an ODE with the predicted velocity.}
 \label{fig:attn_inf}
\end{figure}

The \textbf{ARDiT inference scheme}, including the sequence layout and attention mask during inference, is illustrated in figure~\ref{fig:attn_inf}. Suppose we are generating block $m$. The input to ARDiT is $(C, Z^{<e_{m-1}}, Z_t^{b_m: e_m})$. The attention mask is defined by the following rules: Tokens in $C$ can attend tokens in $C$; tokens in $(Z^{<e_{m-1}}, Z_t^{b_m: e_m})$ can attend $C$ and other tokens with lower or equal block indices.

Let us compare the computational complexity of ARDiTs and decoder-only transformers (LMs) of the same model size. For LMs, we replace the continuous tokens with the same number of discrete tokens. During training, an ARDiT processes $N_{\text{phone}} + 2 \cdot N_{\text{latent}}$ tokens per utterance, while an LM handles $N_{\text{phone}} + N_{\text{latent}}$ tokens. During inference, an ARDiT with a KV cache requires approximately $N_{\text{FE}} + 1$ times as many computations as an LM and $N_{\text{FE}} / B$ times as many network evaluations, where $N_{\text{FE}}$ is the average number of function evaluations of the ODE solver.

Given that both the inference computation and the number of network evaluations in ARDiT grow linearly as $N_{\text{FE}}$ increases, it is crucial to reduce $N_{\text{FE}}$ for practical application. We apply Distribution Matching Distillation (DMD)~\cite{DMD} as described in section \ref{sec:flow} to reduce $N_{\text{FE}}$ to $1$. Detailed explanations of DMD training for ARDiT can be found in Appendix \ref{appendix:dmd_ardit}.

In addition, we can apply fill-in-the-middle (FIM) training to support speech editing with ARDiT. This requires a slightly different attention mask and sequence layout during training. The details of FIM training of ARDiTs can be found in Appendix \ref{appendix:fim_ardit}.

\subsection{Position Embeddings and Total Duration Control in ARDiT TTS}

\label{sec:ardit_pos}

Rather than allowing ARDiT to generate speech of arbitrary length, we use position embeddings to inform ARDiT about the total speech duration $N_{\text{latent}}$. 
Our ARDiT models are built upon Rotary Position Embedding (RoPE) \cite{Roformer}. RoPE encodes relative positional information by rotating the key and value vectors in self-attention. Specifically, a rotation matrix $R_n$ for any position index $n \in \R$ is defined as follows:
\begin{equation}
U_\theta := \begin{bmatrix}
\cos \theta & - \sin\theta\\
\sin \theta & \cos \theta
\end{bmatrix},
\quad 
R_n := \begin{bmatrix} 
U_{n\theta_0} & 0 & \cdots & 0 \\ 
0 & U_{n\theta_1} & \cdots & 0 \\ 
\vdots & \vdots & \ddots & \vdots \\ 
0 & 0 & \cdots & U_{n\theta_{d-1}} \\ 
\end{bmatrix} \in \R^{2d \times 2d}.
\end{equation}
Similar to VALL-E \cite{VALLE}, we employ separate position embeddings for phoneme tokens and speech tokens. For phoneme tokens in $C = [C_0, \cdots, C_{N_{\text{phone}} - 1}]$, each token $C_i$ is given the position index $i$. For speech tokens in $(Z, Z_t, Z_{\mathbf t})$, the tokens $Z^i, Z_t^i, Z_{\mathbf{t}}^i$ are assigned a fractional position index of $i \cdot \eta$, where $\eta = N_{\text{phone}} / N_{\text{latent}}$ is the speech rate.
For instance, if we have $(N_{\text{phone}}, N_{\text{latent}})=(10, 20)$, then the speech rate is $\eta=0.5$, and the fractional position index of the $i$-th speech token is $0.5i$.

This design, similar to the one proposed in ParaNet \cite{ParaNet} for non-autoregressive TTS, was found to accelerate training of ARDiT models in our preliminary experiments. It effectively mitigated the issue of generating overly long sequences, a common problem in autoregressive TTS models. During inference, ARDiT requires an estimate of the speech duration $N_{\text{latent}}$, which can be derived from a reference speech as outlined in Appendix~\ref{appendix:duration}.

\section{Experiments}

\subsection{Setup}
\textbf{Datasets}:
We trained ARDiTs on LibriTTS, a multi-speaker, transcribed English speech dataset, which contains approximately 580 hours of recordings from 2,306 speakers in its training set. We utilized Test Set B from UniCATS~\cite{UniCATS} to evaluate the zero-shot TTS performance. This set contains 500 utterances from 37 speakers in the "test-clean" subset. Each speaker in Test Set B is associated with a speech prompt of approximately 3 seconds. For the speech editing evaluation, we utilized Test Set C from UniCATS~\cite{UniCATS}, which is also the test set used in \cite{RetrieverTTS}. Test Set C contains 25 sentences. Following \cite{UniCATS}, we evaluated text-based speech inpainting performance in two settings with different mask durations. In the short setting, the average mask duration is 0.63 seconds. In the long setting, the average mask duration is 2.18 seconds.

\textbf{Baselines}:
We compare with three non-autoregressive models: 
HierSpeech++\footnote{HierSpeech++: \url{https://github.com/sh-lee-prml/HierSpeechpp}}~\cite{HierSpeechPP},
StyleTTS~2\footnote{StyleTTS 2: \url{https://github.com/yl4579/StyleTTS2}}~\cite{StyleTTS2}, and
UniCATS\footnote{UniCATS: \url{https://github.com/X-LANCE/UniCATS-CTX-vec2wav}}~\cite{UniCATS}; and one autoregressive model,
VoiceCraft\footnote{VoiceCraft: \url{https://github.com/jasonppy/VoiceCraft}}~\cite{VoiceCraft}.
We obtained audio samples of StyleTTS 2, VoiceCraft, and HierSpeech++ using their officially released codes and checkpoints. Since the authors of UniCATS did not provide checkpoints, we directly obtained audio samples from them.
For HierSpeech++, we used the \texttt{lt960} checkpoint. For VoiceCraft, we used \texttt{830M\_TTSEnhanced} model for zero-shot TTS, and \texttt{giga830M} model for speech editing.
For all models, we conducted inference at their native sampling rates, then downsampled the generated audios to 16 kHz for evaluation.

\textbf{Model}:
All transformer models utilized in our study are derived from DiT~\cite{DiT}, including the encoder and decoder of the Mel spectrogram autoencoder, the autoregressive discrete transformers, and the distilled ARDiTs. For the spectrogram encoder, the input time in DiT is set to zero. These models share identical architectures, each comprising 12 layers, 16 heads for multi-head attention, an embedding dimension of 1024, a feed-forward layer dimension of 4096, and a dropout rate of 0.1. We used RoPE \cite{Roformer} as the positional embedding for all the models. More details about the neural network architectures can be found in Appendix~\ref{appendix:arch}. We trained and distilled multiple ARDiTs with various block sizes. All the evaluated ARDiT models were trained with the fill-in-the-middle strategy. Detailed training procedures are available in Appendix~\ref{sec:training_details}.
We utilized the pre-trained BigVGAN \footnote{BigVGAN: \url{https://github.com/NVIDIA/BigVGAN}} vocoder \cite{BigVGAN} with 112M parameters, trained on 24kHz speech, for Mel spectrogram to waveform reconstruction.

\textbf{Subjective Evaluation}:
We conducted a MUSHRA (Multiple Stimuli with Hidden Reference and Anchor) test to evaluate speech naturalness (NAT) and speaker similarity (SIM). We did not use hidden anchors in the test. 20 participants rated audio quality on a scale of 0 to 100. Please refer to Appendix~\ref{sec:mushra} for more details of our listening tests. We report the average MUSHRA scores with 95\% confidence intervals. Additionally, we performed a t-test to evaluate the significance of improvements in the proposed systems.

\textbf{Objective Evaluation}:
We assessed the intelligibility of samples using the Word Error Rate (WER) from the Whisper (medium) ASR model\footnote{Whisper: \url{https://github.com/openai/whisper}}~\cite{Whisper}.
We also measured speaker similarity of generated samples with their prompts using the Speaker Encoding Cosine Similarity (SECS), which ranges from -1 to 1, with higher values indicating greater similarity.
This evaluation utilized three state-of-the-art speaker verification models: WavLM Base with X-vector (WavLM)\footnote{WavLM Base with X-vector: \url{https://huggingface.co/microsoft/wavlm-base-plus-sv}}~\cite{WavLM}, WeSpeaker (WS)\footnote{WeSpeaker: \url{https://github.com/wenet-e2e/wespeaker}}~\cite{WeSpeaker}, and Resemblyzer (Resem)\footnote{Resemblyzer: \url{https://github.com/resemble-ai/Resemblyzer}}~\cite{Resemblyzer}.

\subsection{Zero-Shot Text-to-Speech}\label{sec:result_tts}

In the zero-shot TTS evaluation, we generated speech samples with ARDiTs using the fill-in-the-middle strategy. For each sentence, the prompt speech was given to ARDiTs as both the prefix and suffix, resulting in it being repeated twice. The target total duration was estimated using the prompt speech and prompt text. We applied the Euler sampler with a fixed step size and 16 sampling steps for ODE sampling.

Results in Table~\ref{table:tts} indicate that, in subjective evaluations, student models (DMD) outperform both baselines and teacher models in both metrics. Objectively, the proposed models surpassed baseline models in speaker similarity and achieved comparable WER scores.

\begin{table}[!h]
    \centering
	\caption{Results of zeroshot TTS. For the MUSHRA scores, we performed a t-test comparing ARDiTs with the best baseline model (underlined).\\}
    \label{table:tts}
    \begin{tabular}{ l >{\centering}p{1.6cm} >{\centering}p{1.6cm} c c c c }
        \toprule
        \multicolumn{1}{c}{} & \multicolumn{2}{c}{MUSHRA ($\uparrow$)} & \multicolumn{1}{c}{WER ($\downarrow$)} & \multicolumn{3}{c}{SECS ($\uparrow$)}\\
        \cmidrule(r){2-3} \cmidrule(r){4-4} \cmidrule(r){5-7}
        Model Name                       & NAT                      & SIM                         & Whisper       & WavLM          & WS             & Resem          \\
        \midrule
        Ground Truth                     & 81.1\ci{2.7}             & 52.7\ci{4.9}                & 2.02          & 0.942          & 0.723          & 0.834          \\
        Reconstruct                      & ---                      & ---                         & 2.39          & 0.942          & 0.718          & 0.844          \\
        \midrule
        HierSpeech++~\cite{HierSpeechPP} & 56.1\ci{2.7}             & \underline{59.1\ci{3.7}}    & 5.61          & 0.919          & 0.602          & 0.881          \\
        StyleTTS2~\cite{StyleTTS2}       & \underline{77.2\ci{2.8}} & 52.4\ci{4.1}                & \textbf{1.76} & 0.914          & 0.498          & 0.845          \\
        UniCATS~\cite{UniCATS}           & 53.2\ci{2.8}             & 41.9\ci{3.8}                & 6.33          & 0.912          & 0.537          & 0.832          \\
        VoiceCraft~\cite{VoiceCraft}     & 62.1\ci{3.4}             & 44.9\ci{4.1}                & 4.02          & 0.933          & 0.561          & 0.859          \\
        \midrule
        ARDiT(B=1)                       & 72.7\ci{2.7}             & \px 64.0\ci{4.1}            & 1.83          & \textbf{0.945} & \textbf{0.712} & \textbf{0.886} \\
        ARDiT(B=4)                       & 71.4\ci{2.9}             & 61.4\ci{4.1}                & 2.35          & 0.940          & 0.691          & 0.881          \\
        ARDiT(DMD, B=1)                  & 76.5\ci{3.0}             & \pxxx \textbf{69.4}\ci{3.9} & 1.88          & 0.938          & 0.702          & 0.874          \\
        ARDiT(DMD, B=4)                  & \textbf{79.3}\ci{2.8}    & \pxxx 68.4\ci{4.1}          & 1.81          & 0.933          & 0.656          & 0.867          \\
        \bottomrule\\[-0.8em] \multicolumn{5}{l}{Note: $^{***}p<0.01$, $^{**}p<0.05$, $^{*}p<0.1$}\\
    \end{tabular}\\
\end{table}

\subsection{Speech Editing}

Our speech editing experiment utilized two baseline models, UniCATS and VoiceCraft. We conducted speech editing with ARDiTs using the fill-in-the-middle strategy.
For each test utterance, we provided the ARDiT models with the full text and the remaining audio after masking.
We estimated the target duration using the sections of the audio and text that were not subjected to masking. We applied the Euler sampler with fixed step size and 16 sampling steps for ODE sampling.
In order to minimize the incoherence in the inpainting results, we produced $N_{\text{batch}} = 8$ samples in parallel for each utterance. Subsequently, we employed the post-filtering technique detailed in Appendix Z. This post-filtering process is fully automated and requires no human intervention.
To evaluate the samples, we adopted the MUSHRA test for naturalness and the Word Error Rate (WER) for measuring intelligibility.

The results, detailed in Table~\ref{table:speech_editing}, show that ARDiTs, when distilled with DMD, generally outperform other examined models. Furthermore, all ARDiT models significantly surpassed the baseline models in terms of perceived speech naturalness.

\begin{table}[!h]
	\centering
	\caption{Results on the speech editing task. For the MUSHRA scores, we performed a t-test comparing ARDiTs with the best baseline model (underlined).\\}
    \label{table:speech_editing}
	\begin{tabular}{ l >{\centering}p{1.6cm} >{\centering}p{1.6cm} >{\centering}p{1.6cm} c c c }
		\toprule
		\multicolumn{1}{c}{} & \multicolumn{3}{c}{MUSHRA (Speech Naturalness, $\uparrow$)} & \multicolumn{3}{c}{Whisper WER ($\downarrow$)}\\
		\cmidrule(r){2-4} \cmidrule(r){5-7}
		Model Name                   & short                       & long                        & all                         & short        & long         & all          \\
		\midrule
		Ground Truth                 & 77.8\ci{3.3}                & 78.5\ci{3.4}                & 78.1\ci{2.3}                & 1.33         & 1.33         & 1.33         \\
		\midrule
		UniCATS~\cite{UniCATS}       & \underline{67.0\ci{3.4}}    & \underline{65.5\ci{3.4}}    & \underline{66.2\ci{2.4}}    & 3.05         & 3.18         & 3.11         \\
		VoiceCraft~\cite{VoiceCraft} & 66.9\ci{3.7}                & 62.2\ci{4.6}                & 64.6\ci{2.9}                & 1.59         & 1.99         & 1.79         \\
		\midrule
		ARDiT(R=1)                   & \pxxx 75.5\ci{3.5}          & \pxxx 72.3\ci{3.7}          & \pxxx 73.9\ci{2.5}          & 1.59         & 1.85         & 1.72         \\
		ARDiT(R=4)                   & \pxxx 73.7\ci{3.4}          & \pxx  71.3\ci{3.6}          & \pxxx 72.6\ci{2.4}          & \textbf{1.46}& 1.99         & 1.72         \\
		ARDiT(DMD, R=1)              & \pxxx \textbf{78.1}\ci{3.3} & \pxxx \textbf{77.5}\ci{3.5} & \pxxx \textbf{77.8}\ci{2.4} & 1.85         & \textbf{1.06}& \textbf{1.46}\\
		ARDiT(DMD, R=4)              & \pxxx 75.8\ci{3.4}          & \pxxx 75.2\ci{3.6}          & \pxxx 75.5\ci{2.5}          & 1.46         & 3.31         & 2.38         \\
        \bottomrule\\[-0.8em] \multicolumn{5}{l}{Note: $^{***}p<0.01$, $^{**}p<0.05$, $^{*}p<0.1$}\\
	\end{tabular}
\end{table}

\subsection{Effect of the Block Size} \label{sec:ablation}

\begin{figure}[H]
    \centering
    \includegraphics[width=0.9\textwidth]{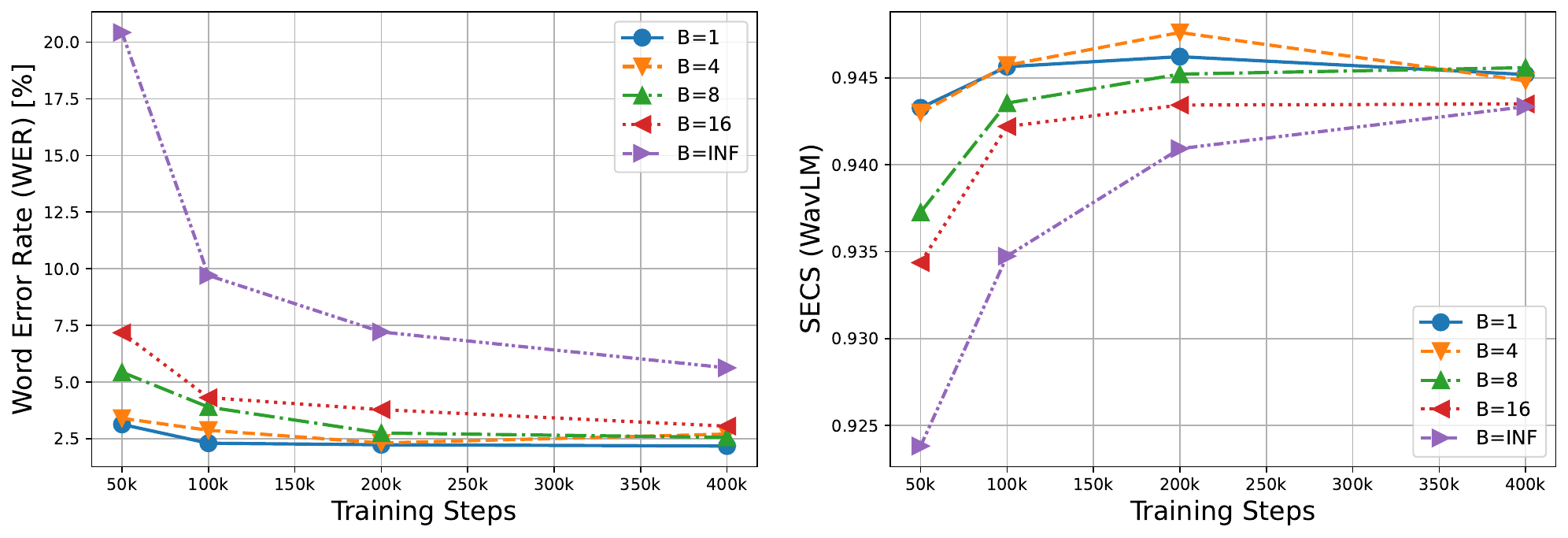}
    \caption{Effects of different block sizes $B$ on ARDiT training. Training efficiency decreases as block size $B$ increases. $B = \text{INF}$ means the block size is infinite, i.e., the model becomes a non-autoregressive (NAR) diffusion model, generating all speech tokens in parallel.}
    \label{fig:ablation}
\end{figure}

Recent works \cite{E3TTS, SimpleTTS, Mapache} in TTS demonstrates that non-autoregressive diffusion models can perform text-to-speech synthesis without force-alignment or explicit phoneme duration modeling. This calls into question the necessity of autoregressive diffusion modeling in TTS.

To answer this question, we trained several ARDiT TTS models under the same settings, only altering the block size $B$ and observed its impact on performance. We examined 5 different settings $B \in \cb{1, 4, 8, 16, \text{INF}}$, where $B=\text{INF}$ represents generating all frames at once, thus serving as a non-autoregressive model. We generated 500 samples from each model as in the zero-shot TTS evaluation win section \ref{sec:result_tts}. For all models we apply Euler sampler with 16 sampling steps, in order to fix the amount of total computation in inference. We evaluated the intelligibility and speaker similarity of the outputs using Word Error Rate (WER) and Speaker Encoding Cosine Similarity (SECS), respectively. Our model's sampling rate is 24kHz, and the tokenization hop size is 1024, resulting in block lengths of approximately 171ms for $B=4$ and 680ms for $B=16$, for example. The average length of our test samples is 5.76 secs, meaning that, in $B=16$, the model generates 8.47 blocks on average to complete the utterance.

Figure~\ref{fig:ablation} illustrates WER and SECS scores under various conditions. Our findings indicate a decrease in WER with increasing training steps and lower $B$ values. Notably, the autoregressive models achieved much better speech intelligibility compared to the non-autoregressive model ($B=\text{INF}$). SECS scores followed the same pattern as WER, showing enhanced speaker similarity with more training and lower block size $B$.

\section{Discussion and Conclusion}

The goal of this research is to eliminate the need for discrete tokenization of audio in language modeling. We proposed the use of continuous tokens in place of discrete ones and applied autoregressive diffusion Transformers (ARDiTs) to generate these continuous audio tokens. Our results demonstrated that ARDiTs perform well in zero-shot text-to-speech and text-based speech editing. Through distillation, an ARDiT TTS model can generate one or more continuous speech tokens in a single network evaluation while maintaining high sample quality.

\noindent
\textbf{Limitations}
Due to resource constraints, we trained and evaluated ARDiTs solely on LibriTTS, which contains only reading-style speech from English audiobooks. We plan to further investigate the applicability of ARDiTs to more diverse data, such as large-scale in-the-wild datasets, in future work.

We only tuned ARDiTs using a limited set of hyperparameters due to resource constraints. Thus, we cannot guarantee the general applicability of our findings in different settings.

We only tested ARDiTs on zero-shot text-to-speech and text-based speech editing; we do not guarantee high performance for more extensive audio generation tasks. We plan to apply ARDiTs to other audio generative tasks in future work.

We trained ARDiTs exclusively for audio generation. It remains uncertain whether the model can be trained to generate both discrete and continuous tokens, similar to existing Multimodal Large Language Models.

It's well acknowledged that the sample quality of language models can be enhanced with annealed sampling techniques like beam search. Such techniques are not currently available for ARDiTs. We intend to research ARDiT sampling techniques in our future work.

\bibliographystyle{IEEEtran}
\bibliography{refs}

\begin{thebibliography}{100}
\providecommand{\url}[1]{#1}
\csname url@samestyle\endcsname
\providecommand{\newblock}{\relax}
\providecommand{\bibinfo}[2]{#2}
\providecommand{\BIBentrySTDinterwordspacing}{\spaceskip=0pt\relax}
\providecommand{\BIBentryALTinterwordstretchfactor}{4}
\providecommand{\BIBentryALTinterwordspacing}{\spaceskip=\fontdimen2\font plus
\BIBentryALTinterwordstretchfactor\fontdimen3\font minus
  \fontdimen4\font\relax}
\providecommand{\BIBforeignlanguage}[2]{{%
\expandafter\ifx\csname l@#1\endcsname\relax
\typeout{** WARNING: IEEEtran.bst: No hyphenation pattern has been}%
\typeout{** loaded for the language `#1'. Using the pattern for}%
\typeout{** the default language instead.}%
\else
\language=\csname l@#1\endcsname
\fi
#2}}
\providecommand{\BIBdecl}{\relax}
\BIBdecl

\bibitem{AudioLMSurvey}
H.~Wu, X.~Chen, Y.-C. Lin, K.-W. Chang, H.-L. Chung, A.~H. Liu, and H.~yi~Lee,
  ``Towards audio language modeling - an overview,'' \emph{arXiv preprint
  arXiv:2402.13236}, 2024.

\bibitem{AudioLM}
Z.~Borsos, R.~Marinier, D.~Vincent, E.~Kharitonov, O.~Pietquin, M.~Sharifi,
  D.~Roblek, O.~Teboul, D.~Grangier, M.~Tagliasacchi \emph{et~al.},
  ``{AudioLM}: a language modeling approach to audio generation,''
  \emph{TASLP}, 2023.

\bibitem{MusicLM}
A.~Agostinelli, T.~I. Denk, Z.~Borsos, J.~Engel, M.~Verzetti, A.~Caillon,
  Q.~Huang, A.~Jansen, A.~Roberts, M.~Tagliasacchi \emph{et~al.}, ``{MusicLM}:
  Generating music from text,'' \emph{arXiv preprint arXiv:2301.11325}, 2023.

\bibitem{VALLE}
C.~Wang, S.~Chen, Y.~Wu, Z.~Zhang, L.~Zhou, S.~Liu, Z.~Chen, Y.~Liu, H.~Wang,
  J.~Li \emph{et~al.}, ``Neural codec language models are zero-shot text to
  speech synthesizers,'' \emph{arXiv preprint arXiv:2301.02111}, 2023.

\bibitem{AudioGen}
F.~Kreuk, G.~Synnaeve, A.~Polyak, U.~Singer, A.~D{\'e}fossez, J.~Copet,
  D.~Parikh, Y.~Taigman, and Y.~Adi, ``{AudioGen}: Textually guided audio
  generation,'' \emph{arXiv preprint arXiv:2209.15352}, 2022.

\bibitem{VIOLA}
T.~Wang, L.~Zhou, Z.~Zhang, Y.~Wu, S.~Liu, Y.~Gaur, Z.~Chen, J.~Li, and F.~Wei,
  ``Vio{LA}: Unified codec language models for speech recognition, synthesis,
  and translation,'' \emph{arXiv preprint arXiv:2305.16107}, 2023.

\bibitem{MusicGen}
J.~Copet, F.~Kreuk, I.~Gat, T.~Remez, D.~Kant, G.~Synnaeve, Y.~Adi, and
  A.~Defossez, ``Simple and controllable music generation,'' in \emph{NeurIPS},
  2023.

\bibitem{AudioPaLM}
P.~K. Rubenstein, C.~Asawaroengchai, D.~D. Nguyen, A.~Bapna, Z.~Borsos,
  F.~d.~C. Quitry, P.~Chen, D.~E. Badawy, W.~Han, E.~Kharitonov \emph{et~al.},
  ``{AudioPaLM}: A large language model that can speak and listen,''
  \emph{arXiv preprint arXiv:2306.12925}, 2023.

\bibitem{SpeechX}
X.~Wang, M.~Thakker, Z.~Chen, N.~Kanda, S.~E. Eskimez, S.~Chen, M.~Tang,
  S.~Liu, J.~Li, and T.~Yoshioka, ``{SpeechX}: Neural codec language model as a
  versatile speech transformer,'' \emph{arXiv preprint arXiv:2308.06873}, 2023.

\bibitem{LauraGPT}
Q.~Chen, Y.~Chu, Z.~Gao, Z.~Li, K.~Hu, X.~Zhou, J.~Xu, Z.~Ma, W.~Wang, S.~Zheng
  \emph{et~al.}, ``{LauraGPT}: Listen, attend, understand, and regenerate audio
  with gpt,'' \emph{arXiv preprint arXiv:2310.04673}, 2023.

\bibitem{UniAudio}
D.~Yang, J.~Tian, X.~Tan, R.~Huang, S.~Liu, X.~Chang, J.~Shi, S.~Zhao, J.~Bian,
  X.~Wu \emph{et~al.}, ``{UniAudio}: An audio foundation model toward universal
  audio generation,'' \emph{arXiv preprint arXiv:2310.00704}, 2023.

\bibitem{GSLM}
K.~Lakhotia, E.~Kharitonov, W.-N. Hsu, Y.~Adi, A.~Polyak, B.~Bolte, T.-A.
  Nguyen, J.~Copet, A.~Baevski, A.~Mohamed \emph{et~al.}, ``On generative
  spoken language modeling from raw audio,'' \emph{TACL}, 2021.

\bibitem{TWIST}
M.~Hassid, T.~Remez, T.~A. Nguyen, I.~Gat, A.~Conneau, F.~Kreuk, J.~Copet,
  A.~Defossez, G.~Synnaeve, E.~Dupoux \emph{et~al.}, ``Textually pretrained
  speech language models,'' \emph{NeurIPS}, 2024.

\bibitem{UnitY}
H.~Inaguma, S.~Popuri, I.~Kulikov, P.-J. Chen, C.~Wang, Y.-A. Chung, Y.~Tang,
  A.~Lee, S.~Watanabe, and J.~Pino, ``Unity: Two-pass direct speech-to-speech
  translation with discrete units,'' \emph{arXiv preprint arXiv:2212.08055},
  2022.

\bibitem{SPEARTTS}
E.~Kharitonov, D.~Vincent, Z.~Borsos, R.~Marinier, S.~Girgin, O.~Pietquin,
  M.~Sharifi, M.~Tagliasacchi, and N.~Zeghidour, ``Speak, read and prompt:
  High-fidelity text-to-speech with minimal supervision,'' \emph{TACL}, 2023.

\bibitem{SoundStream}
N.~Zeghidour, A.~Luebs, A.~Omran, J.~Skoglund, and M.~Tagliasacchi,
  ``{SoundStream}: An end-to-end neural audio codec,'' \emph{TASLP}, 2021.

\bibitem{EnCodec}
A.~D{\'e}fossez, J.~Copet, G.~Synnaeve, and Y.~Adi, ``High fidelity neural
  audio compression,'' \emph{TMLR}, 2023.

\bibitem{AudioDec}
Y.-C. Wu, I.~D. Gebru, D.~Markovi{\'c}, and A.~Richard, ``{AudioDec}: An
  open-source streaming high-fidelity neural audio codec,'' in \emph{ICASSP},
  2023.

\bibitem{HifiCodec}
D.~Yang, S.~Liu, R.~Huang, J.~Tian, C.~Weng, and Y.~Zou, ``{HiFi}-{C}odec:
  Group-residual vector quantization for high fidelity audio codec,''
  \emph{arXiv preprint arXiv:2305.02765}, 2023.

\bibitem{DAC}
R.~Kumar, P.~Seetharaman, A.~Luebs, I.~Kumar, and K.~Kumar, ``High-fidelity
  audio compression with improved rvqgan,'' \emph{NeurIPS}, 2024.

\bibitem{SpeechTokenizer}
X.~Zhang, D.~Zhang, S.~Li, Y.~Zhou, and X.~Qiu, ``{SpeechTokenizer}: Unified
  speech tokenizer for speech large language models,'' \emph{arXiv preprint
  arXiv:2308.16692}, 2023.

\bibitem{FunCodec}
Z.~Du, S.~Zhang, K.~Hu, and S.~Zheng, ``{FunCodec}: A fundamental, reproducible
  and integrable open-source toolkit for neural speech codec,'' in
  \emph{ICASSP}, 2024.

\bibitem{RDPTradeoff}
Y.~Blau and T.~Michaeli, ``Rethinking lossy compression: The
  rate-distortion-perception tradeoff,'' in \emph{ICML}, 2019.

\bibitem{PDTradeoff}
------, ``The perception-distortion tradeoff,'' in \emph{CVPR}, 2018.

\bibitem{VoiceCraft}
P.~Peng, P.-Y. Huang, A.~Mohamed, and D.~Harwath, ``{VoiceCraft}: Zero-shot
  speech editing and text-to-speech in the wild,'' \emph{arXiv preprint
  arXiv:2403.16973}, 2024.

\bibitem{FSQ}
F.~Mentzer, D.~Minnen, E.~Agustsson, and M.~Tschannen, ``Finite scalar
  quantization: {VQ-VAE} made simple,'' \emph{arXiv preprint arXiv:2309.15505},
  2023.

\bibitem{DiscreteResynthesis}
A.~Polyak, Y.~Adi, J.~Copet, E.~Kharitonov, K.~Lakhotia, W.-N. Hsu, A.~Mohamed,
  and E.~Dupoux, ``Speech resynthesis from discrete disentangled
  self-supervised representations,'' \emph{arXiv preprint arXiv:2104.00355},
  2021.

\bibitem{VAE}
D.~P. Kingma and M.~Welling, ``Auto-encoding variational bayes,'' \emph{arXiv
  preprint arXiv:1312.6114}, 2013.

\bibitem{VAEGAN}
A.~B.~L. Larsen, S.~K. S{\o}nderby, H.~Larochelle, and O.~Winther,
  ``Autoencoding beyond pixels using a learned similarity metric,'' in
  \emph{ICML}, 2016.

\bibitem{VQGAN}
P.~Esser, R.~Rombach, and B.~Ommer, ``Taming transformers for high-resolution
  image synthesis,'' in \emph{CVPR}, 2021.

\bibitem{VQVAE}
A.~Van Den~Oord, O.~Vinyals \emph{et~al.}, ``Neural discrete representation
  learning,'' \emph{NeurIPS}, 2017.

\bibitem{Wav2Vec2}
A.~Baevski, Y.~Zhou, A.~Mohamed, and M.~Auli, ``wav2vec 2.0: A framework for
  self-supervised learning of speech representations,'' \emph{NeurIPS}, 2020.

\bibitem{StraightThrough}
M.~Huh, B.~Cheung, P.~Agrawal, and P.~Isola, ``Straightening out the
  straight-through estimator: Overcoming optimization challenges in vector
  quantized networks,'' in \emph{ICML}, 2023.

\bibitem{GumbelSoftmax}
E.~Jang, S.~Gu, and B.~Poole, ``Categorical reparameterization with
  gumbel-softmax,'' \emph{arXiv preprint arXiv:1611.01144}, 2016.

\bibitem{VQWav2Vec}
A.~Baevski, S.~Schneider, and M.~Auli, ``vq-wav2vec: Self-supervised learning
  of discrete speech representations,'' in \emph{ICLR}, 2019.

\bibitem{GSLMCont}
R.~Algayres, Y.~Adi, T.~A. Nguyen, J.~Copet, G.~Synnaeve, B.~Sagot, and
  E.~Dupoux, ``Generative spoken language model based on continuous word-sized
  audio tokens,'' \emph{arXiv preprint arXiv:2310.05224}, 2023.

\bibitem{NormalizingFlow}
G.~Papamakarios, E.~Nalisnick, D.~J. Rezende, S.~Mohamed, and
  B.~Lakshminarayanan, ``Normalizing flows for probabilistic modeling and
  inference,'' \emph{JMLR}, 2021.

\bibitem{Glow}
D.~P. Kingma and P.~Dhariwal, ``Glow: Generative flow with invertible 1x1
  convolutions,'' \emph{NeurIPS}, 2018.

\bibitem{WaveTacotron}
R.~J. Weiss, R.~Skerry-Ryan, E.~Battenberg, S.~Mariooryad, and D.~P. Kingma,
  ``{Wave-Tacotron}: Spectrogram-free end-to-end text-to-speech synthesis,'' in
  \emph{ICASSP}, 2021.

\bibitem{Flowtron}
R.~Valle, K.~Shih, R.~Prenger, and B.~Catanzaro, ``Flowtron: An autoregressive
  flow-based generative network for text-to-speech synthesis,'' \emph{arXiv
  preprint arXiv:2005.05957}, 2020.

\bibitem{WaveFlow}
W.~Ping, K.~Peng, K.~Zhao, and Z.~Song, ``{WaveFlow}: A compact flow-based
  model for raw audio,'' in \emph{ICML}, 2020.

\bibitem{MDN}
C.~M. Bishop, ``Mixture density networks,'' 1994.

\bibitem{MDNTTSDu}
C.~Du and K.~Yu, ``Phone-level prosody modelling with gmm-based mdn for diverse
  and controllable speech synthesis,'' \emph{TASLP}, 2021.

\bibitem{MDNTTSWang}
X.~Wang, S.~Takaki, and J.~Yamagishi, ``An autoregressive recurrent mixture
  density network for parametric speech synthesis,'' \emph{ICASSP}, 2017.

\bibitem{CLaMTTS}
J.~Kim, K.~Lee, S.~Chung, and J.~Cho, ``{CLaM}-{TTS}: Improving neural codec
  language model for zero-shot text-to-speech,'' \emph{arXiv preprint
  arXiv:2404.02781}, 2024.

\bibitem{PixelCNN++}
T.~Salimans, A.~Karpathy, X.~Chen, and D.~P. Kingma, ``{PixelCNN++}: Improving
  the pixelcnn with discretized logistic mixture likelihood and other
  modifications,'' \emph{arXiv preprint arXiv:1701.05517}, 2017.

\bibitem{ParallelWaveNet}
A.~Oord, Y.~Li, I.~Babuschkin, K.~Simonyan, O.~Vinyals, K.~Kavukcuoglu,
  G.~Driessche, E.~Lockhart, L.~Cobo, F.~Stimberg \emph{et~al.}, ``{Parallel
  Wavenet}: Fast high-fidelity speech synthesis,'' in \emph{ICML}, 2018.

\bibitem{GIVT}
M.~Tschannen, C.~Eastwood, and F.~Mentzer, ``{GIVT}: Generative
  infinite-vocabulary transformers,'' \emph{arXiv preprint arXiv:2312.02116},
  2023.

\bibitem{Tacotron}
Y.~Wang, R.~Skerry-Ryan, D.~Stanton, Y.~Wu, R.~J. Weiss, N.~Jaitly, Z.~Yang,
  Y.~Xiao, Z.~Chen, S.~Bengio \emph{et~al.}, ``Tacotron: Towards end-to-end
  speech synthesis,'' \emph{arXiv preprint arXiv:1703.10135}, 2017.

\bibitem{GAN}
I.~Goodfellow, J.~Pouget-Abadie, M.~Mirza, B.~Xu, D.~Warde-Farley, S.~Ozair,
  A.~Courville, and Y.~Bengio, ``Generative adversarial networks,''
  \emph{Communications of the ACM}, 2020.

\bibitem{ChunkWaveGAN}
M.~Morrison, R.~Kumar, K.~Kumar, P.~Seetharaman, A.~Courville, and Y.~Bengio,
  ``Chunked autoregressive gan for conditional waveform synthesis,''
  \emph{arXiv preprint arXiv:2110.10139}, 2021.

\bibitem{DPMZero}
J.~Sohl-Dickstein, E.~Weiss, N.~Maheswaranathan, and S.~Ganguli, ``Deep
  unsupervised learning using nonequilibrium thermodynamics,'' in \emph{ICML},
  2015.

\bibitem{DDPM}
J.~Ho, A.~Jain, and P.~Abbeel, ``Denoising diffusion probabilistic models,''
  \emph{NeurIPS}, 2020.

\bibitem{ScoreSDE}
Y.~Song, J.~Sohl-Dickstein, D.~P. Kingma, A.~Kumar, S.~Ermon, and B.~Poole,
  ``Score-based generative modeling through stochastic differential
  equations,'' \emph{arXiv preprint arXiv:2011.13456}, 2020.

\bibitem{SSDLM}
X.~Han, S.~Kumar, and Y.~Tsvetkov, ``{SSD}-{LM}: Semi-autoregressive
  simplex-based diffusion language model for text generation and modular
  control,'' in \emph{ACL}, 2023.

\bibitem{SSDLM2}
X.~Han, S.~Kumar, Y.~Tsvetkov, and M.~Ghazvininejad, ``{SSD}-2: Scaling and
  inference-time fusion of diffusion language models,'' \emph{arXiv preprint
  arXiv:2305.14771}, 2023.

\bibitem{GroupwiseDiffusion}
S.~Lee, G.~Lee, H.~Kim, J.~Kim, and Y.~Uh, ``Sequential data generation with
  groupwise diffusion process,'' \emph{arXiv preprint arXiv:2310.01400}, 2023.

\bibitem{HierMusic}
Z.~Wang, L.~Min, and G.~Xia, ``Whole-song hierarchical generation of symbolic
  music using cascaded diffusion models,'' \emph{arXiv preprint
  arXiv:2405.09901}, 2024.

\bibitem{Jen1}
P.~Li, B.~Chen, Y.~Yao, Y.~Wang, A.~Wang, and A.~Wang, ``{Jen-1}: Text-guided
  universal music generation with omnidirectional diffusion models,''
  \emph{arXiv preprint arXiv:2308.04729}, 2023.

\bibitem{VASA1}
S.~Xu, G.~Chen, Y.-X. Guo, J.~Yang, C.~Li, Z.~Zang, Y.~Zhang, X.~Tong, and
  B.~Guo, ``{VASA-1}: Lifelike audio-driven talking faces generated in real
  time,'' \emph{arXiv preprint arXiv:2404.10667}, 2024.

\bibitem{DiffAR}
R.~Benita, M.~Elad, and J.~Keshet, ``{DiffAR}: Denoising diffusion
  autoregressive model for raw speech waveform generation,'' \emph{arXiv
  preprint arXiv:2310.01381}, 2023.

\bibitem{LAVIE}
Y.~Wang, X.~Chen, X.~Ma, S.~Zhou, Z.~Huang, Y.~Wang, C.~Yang, Y.~He, J.~Yu,
  P.~Yang \emph{et~al.}, ``{LAVIE}: High-quality video generation with cascaded
  latent diffusion models,'' \emph{arXiv preprint arXiv:2309.15103}, 2023.

\bibitem{VDT}
H.~Lu, G.~Yang, N.~Fei, Y.~Huo, Z.~Lu, P.~Luo, and M.~Ding, ``{VDT}:
  General-purpose video diffusion transformers via mask modeling,'' in
  \emph{ICLR}, 2023.

\bibitem{DiffInstruct}
W.~Luo, T.~Hu, S.~Zhang, J.~Sun, Z.~Li, and Z.~Zhang, ``Diff-{I}nstruct: A
  universal approach for transferring knowledge from pre-trained diffusion
  models,'' \emph{NeurIPS}, 2024.

\bibitem{ScoreGAN}
J.-Y. Franceschi, M.~Gartrell, L.~Dos~Santos, T.~Issenhuth, E.~de~B{\'e}zenac,
  M.~Chen, and A.~Rakotomamonjy, ``Unifying {GAN}s and score-based diffusion as
  generative particle models,'' \emph{NeurIPS}, 2024.

\bibitem{VSD}
Z.~Wang, C.~Lu, Y.~Wang, F.~Bao, C.~Li, H.~Su, and J.~Zhu, ``{ProlificDreamer}:
  High-fidelity and diverse text-to-3d generation with variational score
  distillation,'' \emph{NeurIPS}, 2024.

\bibitem{DMD}
T.~Yin, M.~Gharbi, R.~Zhang, E.~Shechtman, F.~Durand, W.~T. Freeman, and
  T.~Park, ``One-step diffusion with distribution matching distillation,''
  \emph{arXiv preprint arXiv:2311.18828}, 2023.

\bibitem{SiD}
M.~Zhou, H.~Zheng, Z.~Wang, M.~Yin, and H.~Huang, ``{S}core identity
  distillation: Exponentially fast distillation of pretrained diffusion models
  for one-step generation,'' \emph{arXiv preprint arXiv:2404.04057}, 2024.

\bibitem{FIM}
M.~Bavarian, H.~Jun, N.~Tezak, J.~Schulman, C.~McLeavey, J.~Tworek, and
  M.~Chen, ``Efficient training of language models to fill in the middle,''
  \emph{arXiv preprint arXiv:2207.14255}, 2022.

\bibitem{Roformer}
J.~Su, M.~Ahmed, Y.~Lu, S.~Pan, W.~Bo, and Y.~Liu, ``Roformer: Enhanced
  transformer with rotary position embedding,'' \emph{Neurocomputing}, 2024.

\bibitem{VideoDiffusionModels}
J.~Ho, T.~Salimans, A.~Gritsenko, W.~Chan, M.~Norouzi, and D.~J. Fleet, ``Video
  diffusion models,'' in \emph{NeurIPS}, 2022.

\bibitem{RollingDiffusionModels}
D.~Ruhe, J.~Heek, T.~Salimans, and E.~Hoogeboom, ``Rolling diffusion models,''
  \emph{arXiv preprint arXiv:2402.09470}, 2024.

\bibitem{TacotronSpeaker}
Y.~Jia, Y.~Zhang, R.~Weiss, Q.~Wang, J.~Shen, F.~Ren, P.~Nguyen, R.~Pang,
  I.~Lopez~Moreno, Y.~Wu \emph{et~al.}, ``Transfer learning from speaker
  verification to multispeaker text-to-speech synthesis,'' \emph{NeurIPS},
  2018.

\bibitem{VALLEX}
Z.~Zhang, L.~Zhou, C.~Wang, S.~Chen, Y.~Wu, S.~Liu, Z.~Chen, Y.~Liu, H.~Wang,
  J.~Li \emph{et~al.}, ``Speak foreign languages with your own voice:
  Cross-lingual neural codec language modeling,'' \emph{arXiv preprint
  arXiv:2303.03926}, 2023.

\bibitem{MegaTTS}
Z.~Jiang, Y.~Ren, Z.~Ye, J.~Liu, C.~Zhang, Q.~Yang, S.~Ji, R.~Huang, C.~Wang,
  X.~Yin \emph{et~al.}, ``{Mega-TTS}: Zero-shot text-to-speech at scale with
  intrinsic inductive bias,'' \emph{arXiv preprint arXiv:2306.03509}, 2023.

\bibitem{MegaTTS2}
Z.~Jiang, J.~Liu, Y.~Ren, J.~He, C.~Zhang, Z.~Ye, P.~Wei, C.~Wang, X.~Yin,
  Z.~Ma \emph{et~al.}, ``{Mega-TTS} 2: Zero-shot text-to-speech with arbitrary
  length speech prompts,'' \emph{arXiv preprint arXiv:2307.07218}, 2023.

\bibitem{UniCATS}
C.~Du, Y.~Guo, F.~Shen, Z.~Liu, Z.~Liang, X.~Chen, S.~Wang, H.~Zhang, and
  K.~Yu, ``{UniCATS}: A unified context-aware text-to-speech framework with
  contextual vq-diffusion and vocoding,'' in \emph{AAAI}, 2024.

\bibitem{NaturalSpeech2}
K.~Shen, Z.~Ju, X.~Tan, E.~Liu, Y.~Leng, L.~He, T.~Qin, J.~Bian \emph{et~al.},
  ``{NaturalSpeech} 2: Latent diffusion models are natural and zero-shot speech
  and singing synthesizers,'' in \emph{ICLR}, 2023.

\bibitem{NaturalSpeech3}
Z.~Ju, Y.~Wang, K.~Shen, X.~Tan, D.~Xin, D.~Yang, Y.~Liu, Y.~Leng, K.~Song,
  S.~Tang \emph{et~al.}, ``{NaturalSpeech} 3: Zero-shot speech synthesis with
  factorized codec and diffusion models,'' \emph{arXiv preprint
  arXiv:2403.03100}, 2024.

\bibitem{VoiceBox}
M.~Le, A.~Vyas, B.~Shi, B.~Karrer, L.~Sari, R.~Moritz, M.~Williamson,
  V.~Manohar, Y.~Adi, J.~Mahadeokar \emph{et~al.}, ``{VoiceBox}: Text-guided
  multilingual universal speech generation at scale,'' \emph{NeurIPS}, 2024.

\bibitem{Voco}
Z.~Jin, G.~J. Mysore, S.~Diverdi, J.~Lu, and A.~Finkelstein, ``Voco: Text-based
  insertion and replacement in audio narration,'' \emph{TOG}, 2017.

\bibitem{EditSpeech}
D.~Tan, L.~Deng, Y.~T. Yeung, X.~Jiang, X.~Chen, and T.~Lee, ``{EditSpeech}: A
  text based speech editing system using partial inference and bidirectional
  fusion,'' in \emph{ASRU}, 2021.

\bibitem{CampNet}
T.~Wang, J.~Yi, L.~Deng, R.~Fu, J.~Tao, and Z.~Wen, ``Context-aware mask
  prediction network for end-to-end text-based speech editing,'' in
  \emph{ICASSP}, 2022.

\bibitem{A3T}
H.~Bai, R.~Zheng, J.~Chen, M.~Ma, X.~Li, and L.~Huang, ``{A$^3$T}:
  Alignment-aware acoustic and text pretraining for speech synthesis and
  editing,'' in \emph{ICML}, 2022.

\bibitem{SpeechPainter}
Z.~Borsos, M.~Sharifi, and M.~Tagliasacchi, ``{SpeechPainter}: Text-conditioned
  speech inpainting,'' in \emph{Interspeech}, 2022.

\bibitem{FluentSpeech}
Z.~Jiang, Q.~Yang, J.~Zuo, Z.~Ye, R.~Huang, Y.~Ren, and Z.~Zhao,
  ``{F}luent{S}peech: Stutter-oriented automatic speech editing with
  context-aware diffusion models,'' in \emph{ACL}, 2023.

\bibitem{DiffVoice}
Z.~Liu, Y.~Guo, and K.~Yu, ``{DiffVoice}: Text-to-speech with latent
  diffusion,'' in \emph{ICASSP}, 2023.

\bibitem{EdiTTS}
J.~Tae, H.~Kim, and T.~Kim, ``{EdiTTS}: Score-based editing for controllable
  text-to-speech,'' \emph{arXiv preprint arXiv:2110.02584}, 2021.

\bibitem{RetrieverTTS}
D.~Yin, C.~Tang, Y.~Liu, X.~Wang, Z.~Zhao, Y.~Zhao, Z.~Xiong, S.~Zhao, and
  C.~Luo, ``{RetrieverTTS}: Modeling decomposed factors for text-based speech
  insertion,'' \emph{arXiv preprint arXiv:2206.13865}, 2022.

\bibitem{E3TTS}
Y.~Gao, N.~Morioka, Y.~Zhang, and N.~Chen, ``{E3 TTS}: Easy end-to-end
  diffusion-based text to speech,'' \emph{arXiv preprint arXiv:2311.00945},
  2023.

\bibitem{BlogDistillation}
\BIBentryALTinterwordspacing
S.~Dieleman, ``The paradox of diffusion distillation,'' 2024. [Online].
  Available: \url{https://sander.ai/2024/02/28/paradox.html}
\BIBentrySTDinterwordspacing

\bibitem{DirectDD}
E.~Luhman and T.~Luhman, ``Knowledge distillation in iterative generative
  models for improved sampling speed,'' \emph{arXiv preprint arXiv:2101.02388},
  2021.

\bibitem{ProgressiveDistillation}
T.~Salimans and J.~Ho, ``Progressive distillation for fast sampling of
  diffusion models,'' in \emph{ICLR}, 2021.

\bibitem{ConsistencyModel}
Y.~Song, P.~Dhariwal, M.~Chen, and I.~Sutskever, ``Consistency models,'' in
  \emph{ICML}, 2023.

\bibitem{TRACT}
D.~Berthelot, A.~Autef, J.~Lin, D.~A. Yap, S.~Zhai, S.~Hu, D.~Zheng,
  W.~Talbott, and E.~Gu, ``{TRACT}: Denoising diffusion models with transitive
  closure time-distillation,'' \emph{arXiv preprint arXiv:2303.04248}, 2023.

\bibitem{BOOT}
J.~Gu, S.~Zhai, Y.~Zhang, L.~Liu, and J.~Susskind, ``{BOOT}: Data-free
  distillation of denoising diffusion models with bootstrapping,'' \emph{arXiv
  preprint arXiv:2306.05544}, 2023.

\bibitem{DSNO}
H.~Zheng, W.~Nie, A.~Vahdat, K.~Azizzadenesheli, and A.~Anandkumar, ``Fast
  sampling of diffusion models via operator learning,'' \emph{arXiv preprint
  arXiv:2211.13449}, 2022.

\bibitem{TCD}
J.~Zheng, M.~Hu, Z.~Fan, C.~Wang, C.~Ding, D.~Tao, and T.-J. Cham, ``Trajectory
  consistency distillation,'' \emph{arXiv preprint arXiv:2402.19159}, 2024.

\bibitem{DDIM}
J.~Song, C.~Meng, and S.~Ermon, ``Denoising diffusion implicit models,''
  \emph{arXiv preprint arXiv:2010.02502}, 2020.

\bibitem{ProDiff}
R.~Huang, Z.~Zhao, H.~Liu, J.~Liu, C.~Cui, and Y.~Ren, ``{ProDiff}: Progressive
  fast diffusion model for high-quality text-to-speech,'' in \emph{ACM
  Multimedia}, 2022.

\bibitem{ReflowTTS}
W.~Guan, Q.~Su, H.~Zhou, S.~Miao, X.~Xie, L.~Li, and Q.~Hong, ``{Reflow-TTS}: A
  rectified flow model for high-fidelity text-to-speech,'' in \emph{ICASSP},
  2024.

\bibitem{CoMoSpeech}
Z.~Ye, W.~Xue, X.~Tan, J.~Chen, Q.~Liu, and Y.~Guo, ``{CoMoSpeech}: One-step
  speech and singing voice synthesis via consistency model,'' in \emph{ACM
  Multimedia}, 2023.

\bibitem{DiffGANTTS}
S.~Liu, D.~Su, and D.~Yu, ``{DiffGAN-TTS}: High-fidelity and efficient
  text-to-speech with denoising diffusion gans,'' \emph{arXiv preprint
  arXiv:2201.11972}, 2022.

\bibitem{Reflow}
X.~Liu, C.~Gong \emph{et~al.}, ``Flow straight and fast: Learning to generate
  and transfer data with rectified flow,'' in \emph{NeurIPS Workshop on
  Score-Based Methods}, 2022.

\bibitem{FlowMatching}
Y.~Lipman, R.~T. Chen, H.~Ben-Hamu, M.~Nickel, and M.~Le, ``Flow matching for
  generative modeling,'' \emph{arXiv preprint arXiv:2210.02747}, 2022.

\bibitem{StochasticInterpolants}
M.~S. Albergo, N.~M. Boffi, and E.~Vanden-Eijnden, ``Stochastic interpolants: A
  unifying framework for flows and diffusions,'' \emph{arXiv preprint
  arXiv:2303.08797}, 2023.

\bibitem{SiT}
N.~Ma, M.~Goldstein, M.~S. Albergo, N.~M. Boffi, E.~Vanden-Eijnden, and S.~Xie,
  ``{SiT}: Exploring flow and diffusion-based generative models with scalable
  interpolant transformers,'' \emph{arXiv preprint arXiv:2401.08740}, 2024.

\bibitem{Transformer}
A.~Vaswani, N.~Shazeer, N.~Parmar, J.~Uszkoreit, L.~Jones, A.~N. Gomez,
  {\L}.~Kaiser, and I.~Polosukhin, ``Attention is all you need,''
  \emph{NeurIPS}, 2017.

\bibitem{DiT}
W.~Peebles and S.~Xie, ``Scalable diffusion models with transformers,'' in
  \emph{ICCV}, 2023.

\bibitem{MIBounds}
B.~Poole, S.~Ozair, A.~Van Den~Oord, A.~Alemi, and G.~Tucker, ``On variational
  bounds of mutual information,'' in \emph{ICML}, 2019.

\bibitem{ParaNet}
K.~Peng, W.~Ping, Z.~Song, and K.~Zhao, ``Non-autoregressive neural
  text-to-speech,'' in \emph{ICML}, 2020.

\bibitem{HierSpeechPP}
S.-H. Lee, H.-Y. Choi, S.-B. Kim, and S.-W. Lee, ``{HierSpeech++}: Bridging the
  gap between semantic and acoustic representation of speech by hierarchical
  variational inference for zero-shot speech synthesis,'' \emph{arXiv preprint
  arXiv:2311.12454}, 2023.

\bibitem{StyleTTS2}
Y.~A. Li, C.~Han, V.~Raghavan, G.~Mischler, and N.~Mesgarani, ``Style{TTS} 2:
  Towards human-level text-to-speech through style diffusion and adversarial
  training with large speech language models,'' \emph{NeurIPS}, 2024.

\bibitem{BigVGAN}
S.-g. Lee, W.~Ping, B.~Ginsburg, B.~Catanzaro, and S.~Yoon, ``{BigVGAN}: A
  universal neural vocoder with large-scale training,'' \emph{arXiv preprint
  arXiv:2206.04658}, 2022.

\bibitem{Whisper}
A.~Radford, J.~W. Kim, T.~Xu, G.~Brockman, C.~McLeavey, and I.~Sutskever,
  ``Robust speech recognition via large-scale weak supervision,'' \emph{arXiv
  preprint arXiv:2212.04356}, 2022.

\bibitem{WavLM}
S.~Chen, C.~Wang, Z.~Chen, Y.~Wu, S.~Liu, Z.~Chen, J.~Li, N.~Kanda,
  T.~Yoshioka, X.~Xiao, J.~Wu, L.~Zhou, S.~Ren, Y.~Qian, Y.~Qian, J.~Wu,
  M.~Zeng, and F.~Wei, ``{WavLM}: Large-scale self-supervised pre-training for
  full stack speech processing,'' \emph{arXiv preprint arXiv:2110.13900}, 2021.

\bibitem{WeSpeaker}
H.~Wang, C.~Liang, S.~Wang, Z.~Chen, B.~Zhang, X.~Xiang, Y.~Deng, and Y.~Qian,
  ``{WeSpeaker}: A research and production oriented speaker embedding learning
  toolkit,'' in \emph{ICASSP}, 2023.

\bibitem{Resemblyzer}
L.~Wan, Q.~Wang, A.~Papir, and I.~L. Moreno, ``Generalized end-to-end loss for
  speaker verification,'' 2020.

\bibitem{SimpleTTS}
\BIBentryALTinterwordspacing
J.~Lovelace, S.~Ray, K.~Kim, K.~Q. Weinberger, and F.~Wu, ``{Simple-TTS}:
  End-to-end text-to-speech synthesis with latent diffusion,'' 2023. [Online].
  Available: \url{https://openreview.net/forum?id=m4mwbPjOwb}
\BIBentrySTDinterwordspacing

\bibitem{Mapache}
G.~C{\'a}mbara, P.~L. Tobing, M.~Babianski, R.~Vipperla, D.~W.~R. Shmelkin,
  G.~Coccia, O.~Angelini, A.~Joly, M.~Lajszczak, and V.~Pollet, ``Mapache:
  Masked parallel transformer for advanced speech editing and synthesis,'' in
  \emph{ICASSP}, 2024.

\bibitem{ScoreMatching}
P.~Vincent, ``A connection between score matching and denoising autoencoders,''
  \emph{Neural Computation}, 2011.

\bibitem{Thienpondt_2021}
J.~Thienpondt, B.~Desplanques, and K.~Demuynck, ``The {IDLAB} {VoxSRC-20}
  submission: Large margin fine-tuning and quality-aware score calibration in
  dnn based speaker verification,'' \emph{arXiv preprint arXiv:2010.11255},
  2020.

\bibitem{VLAE}
X.~Chen, D.~P. Kingma, T.~Salimans, Y.~Duan, P.~Dhariwal, J.~Schulman,
  I.~Sutskever, and P.~Abbeel, ``Variational lossy autoencoder,'' \emph{arXiv
  preprint arXiv:1611.02731}, 2016.

\bibitem{BitsBack}
G.~E. Hinton and D.~van Camp, ``Keeping the neural networks simple by
  minimizing the description length of the weights,'' in \emph{Annual
  Conference Computational Learning Theory}, 1993.

\bibitem{AdamW}
I.~Loshchilov and F.~Hutter, ``Decoupled weight decay regularization,'' in
  \emph{ICLR}, 2017.

\bibitem{BeaqleJS}
\BIBentryALTinterwordspacing
S.~Kraft and U.~Z{\"o}lzer, ``{BeaqleJS}: {HTML5} and {JavaScript} based
  framework for the subjective evaluation of audio quality,'' in \emph{Linux
  Audio Conference}, 2014. [Online]. Available:
  \url{https://github.com/HSU-ANT/beaqlejs}
\BIBentrySTDinterwordspacing

\end{thebibliography}

\appendix
\clearpage
\section{Recover Score Functions from Velocity Fields in Flow Matching Models}

\label{appendix:flow_score}

Suppose $X, Z$ are independent $\R^d$-valued random variables with data density $p(x)$ and Gaussian density $p(z) = \nN(0, I_d)$. Let $\alpha_t = (1 - t)$ and $\sigma_t = t$ for $t \in [0, 1]$. Let $X_t = \alpha_t X + \sigma_t Z$. Suppose $X_t \sim p_t(x)$. We can show that in a Rectified Flow \cite{Reflow} model trained on mapping Gaussian noise to $p(x)$, the velocity field $v(x_t, t)$ is a linear combination of $x_t$ and score function $s(x_t, t) = \nabla_{x_t} \log p_t(x_t)$ for each $t \in [0, 1]$.

Define the following functions:
\begin{align}
\epsilon(x_t, t) &:= \argmin_{\epsilon} E {\norm{\epsilon(X_t, t) - Z}_2^2} = E[Z | X_t = x_t];\\
\mu(x_t, t) &:= \argmin_{\mu}E\norm{\mu(X_t, t) - X}_2^2 = E[X|X_t = x_t];\\
s(x_t, t) &:= \argmin_{s} E\norm{s(X_t, t) + {Z}/{\sigma_t}}_2^2 = E[-Z/\sigma_t | X_t = x_t];
\end{align}
From denoising score matching \cite{ScoreMatching}, we have $s(x_t, t)=\nabla_{x_t} \log p_{t}(x_t)$. Also by definition of $X_t$:
\begin{equation}
\mu(x_t, t) = \frac{1}{\alpha_t}\cb{E\bb{\alpha_tX + \sigma_t Z | X_t = x_t} - \sigma_t E\bb{Z | X_t = x_t}} = \frac{x_t - \sigma_t \epsilon(x_t, t)}{\alpha_t}.
\end{equation}
Therefore $v(x_t, t)$ is a linear combination of $x_t$ and $s(x_t, t)$ and $s(x_t, t)$ is a linear combination of $x_t$ and $v(x_t, t)$.
\begin{equation}
v(x_t, t) = \epsilon(x_t, t) - \mu(x_t, t) = -\sigma_t s(x_t, t) - \frac{x_t + \sigma_t^2 s(x_t, t)}{\alpha_t} = \frac{-1}{1 - t}x_t + \frac{-t}{1 - t} s(x_t, t).
\end{equation}

\section{Discussion of Speech Recognition and Speaker Verification Models for Objective Evaluation }

In addition to the objective evaluation results presented in Table~\ref{table:tts}, we utilized additional speech recognition and speaker verification models to assess Speaker Encoding Cosine Similarity (SECS) and Word Error Rate (WER), respectively. The comprehensive results for zero-shot TTS are detailed in Table~\ref{table:complete}. For WER, we expanded our evaluation to include various sizes of the Whisper model; besides the medium-size model with 769M parameters, we incorporated the base model (74M) and the small model (244M). For SECS, we introduced the WeSpeaker with large-margin finetuning (WS-LM)~\cite{Thienpondt_2021}, which enhances robustness.

As a result, WER decreases for larger models, affirming the reliability of whisper's WER. Regarding SECS, the best model (ARDiT($B=1$)) exhibits consistency across speaker verification models. WavLM, WS, and WS-LM demonstrate a comparable pattern, with higher values observed in ground truth and ARDiT, and lower values in baseline models. However, the score of ground truth in Resemblyzer notably diminishes, casting doubt on its robustness and reliability in assessing speaker similarity.

\begin{table}[!h]
\caption{Complete Objective Evaluation Result for Zeroshot TTS}
\label{table:complete}
\centering
\begin{tabular}{lccccccc}
\toprule
\midrule
\multicolumn{1}{c}{} & \multicolumn{3}{c}{Whisper WER ($\downarrow$)} & \multicolumn{4}{c}{SECS ($\uparrow$)}\\
\cmidrule(r){2-4} \cmidrule(r){5-8}
 Model Name & base & small & medium & WavLM & WS & WS-LM & Resem\\
\midrule
Ground Truth &3.12 & 2.19 & 2.02 & 0.942 & 0.723 & 0.708 & 0.834\\
Reconstruct &3.88 & 2.52 & 2.39 & 0.942 & 0.718 & 0.694 & 0.844\\
\midrule
HierSpeech++~\cite{HierSpeechPP} &9.42 & 6.67 & 5.61 & 0.919 & 0.602 & 0.570 & 0.881\\
StyleTTS2~\cite{StyleTTS2} &\textbf{2.30} & \textbf{1.88} & \textbf{1.76} & 0.914 & 0.498 & 0.462 & 0.845\\
UniCATS~\cite{UniCATS} &9.33 & 7.18 & 6.33 & 0.912 & 0.537 & 0.505 & 0.832\\
VoiceCraft~\cite{VoiceCraft} &6.28 & 4.55 & 4.02 & 0.933 & 0.561 & 0.526 & 0.859\\
\midrule
ARDiT(B=1) &3.57 & 2.38 & 1.83 & \textbf{0.945} & \textbf{0.712} & \textbf{0.681} & \textbf{0.886}\\
ARDiT(B=4) &4.00 & 2.72 & 2.35 & 0.940 & 0.691 & 0.659 & 0.881\\
ARDiT(DMD, B=1) &3.03 & 2.06 & 1.88 & 0.938 & 0.702 & 0.672 & 0.874\\
BRDiT(DMD, B=4) &3.01 & 1.99 & 1.81 & 0.933 & 0.656 & 0.624 & 0.867\\
\midrule
\bottomrule
\end{tabular}
\end{table}

\section{Bitrate of Mel Spectrogram Autoencoder}

\label{appendix:bitrate}

We adopted the pre-trained BigVGAN~\cite{BigVGAN} to reconstruct waveforms from Mel spectrograms. In our experiments, $D_{\text{mel}} = 100$ and $N_{\text{sample}} / N_{\text{frame}} = 256$. Therefore, the uncompressed log-Mel spectrograms, stored as 32-bit floats, have a bitrate of 300 kbps. Meanwhile, the uncompressed audios sampled at 24 kHz with a 16-bit depth have a bitrate of 384 kbps.

We report $R_{\text{bit}} = E\bb{\dkl{q_\phi(z | Y)}{p(z)} / T_{\text{audio}}}$ as the theoretical bitrate of our Mel spectrogram autoencoder, where $T_{\text{audio}}$ is the total duration of the audio in seconds and $D_\text{KL}$ is in bits. In our experiments, we set $\beta_{\text{MI}} = 0.035$ and obtained $R_{\text{bit}} = 1.7 \text{ kpbs}$. When the mean and variance of $q(z|Y)$ are stored as 32-bit floating point numbers, the resulting bitrate is 24 kbps. 

$E\bb{\dkl{q_\phi(z|Y)}{p(z)}}$ can be considered the expected number of bits required to encode distribution $q_\phi(z|Y)$ \cite{DDPM, VLAE, BitsBack}. Notably, $R_{\text{bit}}$ is also the bitrate of VQ-based audio codecs \cite{EnCodec} where the encoder $q_\phi(z|y)$ is deterministic and $p(z)$ is the uniform distribution on the codebook.

\section{Fine-tuning the Mel Spectrogram Decoder on Masked Reconstruction}

\label{appendix:mel_inpaint}

In section \ref{sec:melctx}, we mentioned that the Mel spectrogram decoder is fine-tuned on the masked reconstruction of Mel spectrograms. In this section, we describe more details of the fine-tuning process.

For clarity, let us first define the \textit{masked column selection operator}. We start with a binary mask $m \in \{0, 1\}^K$ and define $|m| := \sum_{k}m_k$. For a given matrix $A = \{A_0; \cdots; A_{K-1}\} \in \R^{K \times D}$, we designate $A_m \in \R^{|m|\times D}$ as the matrix that includes the $k$th row from $A$ if, and only if, $m_k = 1$. The complement of the mask $m$ is denoted as $\bar m = 1-m$.

Instead of training with equation \ref{equ:mel_vae}, the autoencoder is fine-tuned with loss:
\begin{multline}
\mathcal L_{\text{mask}}(\phi, \psi) = \beta_{\text{MI}} \cdot E_Y\bb{\dkl{p_\phi(z | Y)}{p(z)}} \\+ E_{t \sim \U[0, 1], W \sim \nN(0, I)}\bb{\norm{v_\psi((1 - t) Y_M + t W_M; t, Z, Y_{\bar M}) - \p{W_M - Y_M}}_2^2}.
\end{multline}

$M = \bb{M_0, \cdots, M_{N_{\text{frame}} - 1}}$ is a stochastic binary mask, it is constructed in the following way. Let ${A \sim \U\cb{0, \cdots, N_{\text{frame} - 1}}}$ be a random drawn index. Then define:
\begin{equation*}
    M_n := 1((A + n) \bmod N_{\text{frame}} < N_{\text{frame}} / 2).
\end{equation*}
This mask randomly masks 50\% of all tokens.

\section{Distribution Matching Distillation for ARDiTs}

\label{appendix:dmd_ardit}

In this section, we describe how to apply distribution matching distillation (DMD) to ARDiTs in more detail. Let us continue the discussion in section \ref{sec:ardit}. Suppose we have already trained an ARDiT $v_\theta(z_t^{i:i+B}; t, c, z^{<i})$ on text-to-speech synthesis. As a conditional Flow Matching model, ARDiT establishes a mapping $f_\theta(w^{i:i+B}; c, z^{<i})$ that maps Gaussian noise to data. Evaluating $f_\theta$ is slow, as it requires simulating the ODE in equation \ref{equ:flowode} with $N_{\text{FE}}$ network evaluations.

DMD training involves the interaction of three models: the generator $g_\xi(w^{i:i+B}; t, c, z^{<i})$, the true velocity field estimator $v_\theta$, and the fake velocity estimator $v_\eta$. All three models share the same network architecture and initial parameters, so $\eta = \xi = \theta$ at intiailization. Suppose $v_\xi$ is a copy of $v_\theta$, $g_\xi$ is defined as:
\begin{equation}
g_\xi(w^{i:i + B}; c, z^{<i}) := w^{i:i+B} - v_\xi(w^{i:i+B}; 1,c, z^{<i}).
\end{equation}

After initialization, we optimize $\xi$ and $\eta$ through repeating Algorithm~\ref{alg:regression_loss},~\ref{alg:ikl}, and~\ref{alg:flow_matching}. In Algorithm~\ref{alg:regression_loss}, we can save training time by evaluating $f_\theta$ on many samples in advance \cite{DMD}. The for loops over block index $m$ in Algorithm~\ref{alg:regression_loss},~\ref{alg:ikl}, and~\ref{alg:flow_matching} can be evaluated in parallel with the ARDiT training scheme described in Section~\ref{sec:ardit}.

\begin{samepage}
\begin{algorithm}
\caption{Optimize the Regression Loss}
\label{alg:regression_loss}
\begin{algorithmic}
    \State \textbf{Requires:} Training set $\mathcal{D}$, model parameters $\theta$ and $\xi$, block size $B$, coefficient $\beta_{\text{reg}}$
    
    \Procedure{MinimizeRegressionLoss}{}
        \State Sample a pair $(c, z)$ from the training set $\mathcal{D}$
        \State Sample Gaussian noise $w$ with the same shape as $z$
        \State Sample block shift $S$ from $\{0, \cdots, B - 1\}$
        \State Partition $z$ into $M$ blocks: $z = [z^{b_0:e_0}; \cdots; z^{b_{M-1}:e_{M-1}}]$
        
        \For {each block $0 \leq m < M$}
            \State Compute $\hat{z}^{b_m:e_m} := f_\theta(w^{b_m:e_m}; c, z^{<b_m})$ by sampling from the ODE in equation \eqref{equ:flowode}
            \State Compute $\tilde{z}^{b_m:e_m} := g_\xi(w^{b_m:e_m}; c, z^{<b_m})$
        \EndFor
        
        \State Compute the L2 regression loss:
        \begin{equation}
            \ell_{\text{reg}} := \beta_{\text{reg}} \cdot \| \hat{z} - \tilde{z} \|_2^2
        \end{equation}
        
        \State Compute the gradient of $\xi$ with respect to $\ell_{\text{reg}}$, and take an optimizer step to update $\xi$
    \EndProcedure
\end{algorithmic}
\end{algorithm}

\begin{algorithm}
\caption{Optimize the Integral KL Divergence}
\label{alg:ikl}
\begin{algorithmic}
    \State \textbf{Requires:} Training set $\mathcal{D}$, model parameters $\xi$, $\eta$, and $\theta$, block size $B$
    
    \Procedure{MinimizeIKLDivergence}{}
        \State Sample a pair $(c, z)$ from the training set $\mathcal{D}$
        \State Sample Gaussian noises $w, \tilde{w}$ with the same shape as $z$
        \State Sample block shift $S$ from $\{0, \cdots, B - 1\}$
        \State Partition $z$ into $M$ blocks: $z = [z^{b_0:e_0}; \cdots; z^{b_{M-1}:e_{M-1}}]$
        
        \For {each block $0 \leq m < M$}
            \State Compute $\tilde{z}^{b_m:e_m} := g_\xi(w^{b_m:e_m}; c, z^{<b_m})$
            \State Sample random time $t_m$ uniformly from $(0, 1]$
            \State Compute $\tilde{z}_{\mathbf{t}}^{b_m:e_m} := (1 - t_m) \tilde{z}^{b_m:e_m} + t_m \tilde{w}^{b_m:e_m}$
            \State Compute $\Delta^{b_m:e_m} := v_{\theta}(\tilde{z}_{\mathbf{t}}^{b_m:e_m}; t_m, c, z^{<b_m}) - v_\eta(\tilde{z}_{\mathbf{t}}^{b_m:e_m}; t_m, c, z^{<b_m})$
        \EndFor
        
        \State Compute the IKL loss:
        \begin{equation}
            \ell_{\text{IKL}} := \| \tilde{z} + \text{sg}(\Delta - \tilde{z}) \|_2^2
        \end{equation}
        
        \State Compute gradient of $\xi$ with respect to $\ell_{\text{IKL}}$, and take an optimizer step to update $\xi$
    \EndProcedure
\end{algorithmic}
\end{algorithm}

\begin{algorithm}
\caption{Optimize the Flow Matching Objective}
\label{alg:flow_matching}
\begin{algorithmic}
    \State \textbf{Requires:} Training set $\mathcal{D}$, model parameters $\xi$ and $\eta$, block size $B$
    
    \Procedure{MinimizeFlowMatchingLoss}{}
        \State Sample a pair $(c, z)$ from the training set $\mathcal{D}$
        \State Sample Gaussian noise $w$ with the same shape as $z$
        \State Partition $z$ into $M$ blocks: $z = [z^{b_0:e_0}; \cdots; z^{b_{M-1}:e_{M-1}}]$
        
        \For {each block $0 \leq m < M$}
            \State Compute $\tilde{z}^{b_m:e_m} := g_\xi(w^{b_m:e_m}; c, z^{<b_m})$
            \State Sample random time $t_m$ uniformly from $[0, 1]$
            \State Compute $\tilde{z}_{\mathbf{t}}^{b_m:e_m} := (1 - t_m) \tilde{z}^{b_m:e_m} + t_m w^{b_m:e_m}$
        \EndFor
        
        \State Compute the Flow Matching loss:
        \begin{equation}
            \ell_{\text{FM}} = \sum_{m=0}^{M-1} \| v_\eta(\tilde{z}_{\mathbf{t}}^{b_m:e_m}, t_m, c, z^{<b_m}) - (w^{b_m:e_m} - \tilde{z}^{b_m:e_m}) \|_2^2
        \end{equation}
        
        \State Compute gradient of $\eta$ with respect to $\ell_{\text{FM}}$, and take an optimizer step to update $\eta$
    \EndProcedure
\end{algorithmic}
\end{algorithm}
\end{samepage}

\section{Fill-in-the-Middle Training of ARDiT for Text-to-Speech}

\label{appendix:fim_ardit}

\subsection{FIM Training and Inference with ARDiT}

In this section, we give further details of fill-in-the-middle (FIM) training of ARDiTs. FIM training allows ARDiTs to conduct speech editing when the text is given. Let us continue the discussion in section \ref{sec:ardit}. During FIM training, we randomly split $Z$ into three chunks: the left context $Z^{<N_{\text{L}}}$ the missing middle part $Z^{N_{\text{L}}:N_{\text{R}}}$, and the right context $Z^{\ge N_{\text{R}}}$, where we define $A:B$ as $[A, B)$. We first sample $N_{\text R} - N_{\text L}$ uniformly from $\U\cb{1, \cdots, N_{\text{frame}}}$, then sample $N_\text L$ uniformly from all possible values. Notice that it is possible that the left context and right context can be empty. The training loss of ARDiTs with FIM is:
\begin{equation}
\mathcal L_{\text{FIM}}(\theta) = E_{i , t \sim \U[0, 1]}\norm{v_\theta\p{Z_t^{i:i+B}; t, C, Z^{<i}, Z^{\ge N_\text R}} - \p{W^{i:i+B} - Z^{i:i+B}}}_2^2.
\end{equation}
where $i$ is an index sampled uniformly from $[N_\text L , N_\text R - B)$.

\begin{figure}[ht]
    \centering
    \includegraphics[width=0.9\textwidth,trim=.8cm .2cm 1.7cm 0cm,clip]{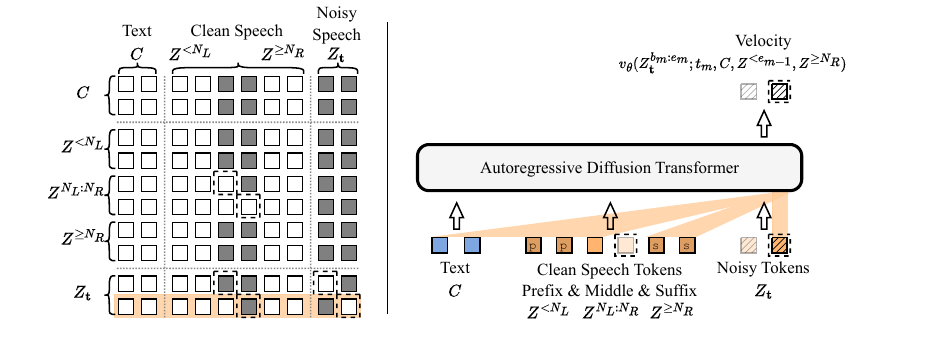}
    \caption{Illustration of the Fill-in-the-middle (FIM) training scheme of ARDiT for text-to-speech. The attention mask is displayed on the left, and the ARDiT model's input and output during FIM training are shown on the right. The prefix, middle, and suffix parts each contain two speech tokens, with a block size of $B = 1$. All tokens are allowed to attend to both prefix and suffix speech tokens.}
    \label{fig:ardit_fim_train}
\end{figure}

The ARDiT training scheme introduced in section \ref{sec:ardit} is still applicable with some modifications. The input sequence in FIM training is now $(C, Z, Z_{\mathbf t}^{N_\text L: N_\text R})$. Tokens in $Z^{N_\text L : N_\text R}$ are chunked into blocks of size $B$ (possibly less than $B$ in the first and last blocks) $Z^{N_\text L: N_\text R} = \bb{Z^{b_0:e_0}; \cdots; Z^{b_{M-1}: e_{M-1}}}$. For each block $m$ there is a random time $t_m \in [0, 1]$. Let $\mathbf t = \bb{t_0, \cdots, t_{M - 1}}$. We define $Z_\mathbf t^{b_m: e_m} := (1 - t_m) Z^{b_m: e_m} + t_m W^{b_m :e_m}$. In FIM training, tokens in $Z^{<N_\text L}$ and $Z^{\ge N_\text R}$ can be attended by all tokens, and they can attend tokens in $C, Z^{<N_{\text L}}$ and $Z^{\ge N_\text R}$. Figure \ref{fig:ardit_fim_train} gives an illustration of the ARDiT FIM training scheme. The position indices of all tokens in $C, Z, Z_{\textbf t}$ are the same as that described in Section \ref{sec:ardit_pos}. Note that we do not need to reorder the sequence $Z, Z_{\mathbf t}$ during training or add sentinel tokens as done in~\cite{FIM}, as the relative position information of the speech tokens is already encoded in their position embeddings.

\begin{figure}[ht]
    \centering
    \includegraphics[width=0.95\textwidth,trim=.9cm .2cm 1.2cm 0cm,clip]{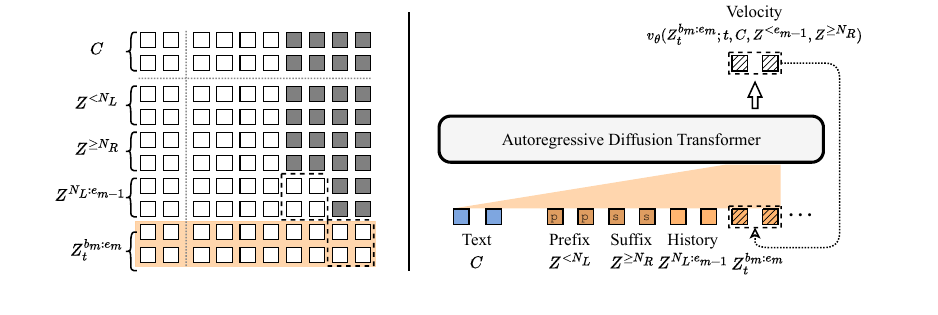}
    \caption{Depiction of the Fill-in-the-middle (FIM) inference scheme of ARDiT for text-to-speech. The attention mask is presented on the left, while the input and output of the ARDiT model during FIM inference are shown on the right. The model has already generated one block of speech tokens and is in the process of generating the second block, with a block size of $B = 2$. The suffix tokens are arranged after prefix tokens to facilitate KV caching during inference.}
    \label{fig:ardit_fim_inference}
\end{figure}

During FIM inference, suppose we are generating block $m$. The input sequence to an ARDiT is $(C, Z^{<N_{\text L}}, Z^{\ge N_\text R}, Z^{N_{\text{L}}:e_{m-1}}, Z_t^{b_m: e_m})$. Here we reordered the sequence to enable autoregressive generation with KV cache. Figure~\ref{fig:ardit_fim_inference} gives an illustration of the ARDiT FIM inference scheme.

\subsection{Post-Filtering for Speech Editing} \label{appendix:edit_post_filter}

In speech editing with ARDiTs, we have noticed occasional failure of the model to smoothly connect generated speech to the suffix. Specifically, we've assessed the probability of detectable failure of model ARDiT (DMD, B=1) on utterances from test set B, where the model was tasked with filling in the middle third of an utterance given the text transcript. The model has an failure rate of approximately 7\%. 

In order to mitigate this issue, we suggest a simple heuristic to evaluate the fluency of the speech generated. Suppose the generated section is $Z^{N_\text{L} : N_\text{R}}$. We then sample the block $Z^{N_{\text R}:N_{\text R} + B}$ with the model, and then calculate the L2 distance of generated $\what Z^{N_{\text R}: N_{\text R} + B}$ to the ground truth. During inference, we generate a batch of $N_{\text{batch}}$ samples from the model and choose the sample with the smallest L2 distance. With this method, we can reduce the failure rate to 2.2\% when $N_{\text{batch}} = 8$. We applied this strategy with $N_{\text{batch}} = 8$ for our speech editing evaluation. We did not apply this strategy for zero-shot TTS evaluations.

\section{Network Architecture Details} \label{appendix:arch}

\begin{figure}[ht]
    \centering
    \includegraphics[width=1.0\textwidth,trim=0.3cm 0cm 0.2cm 0cm,clip]{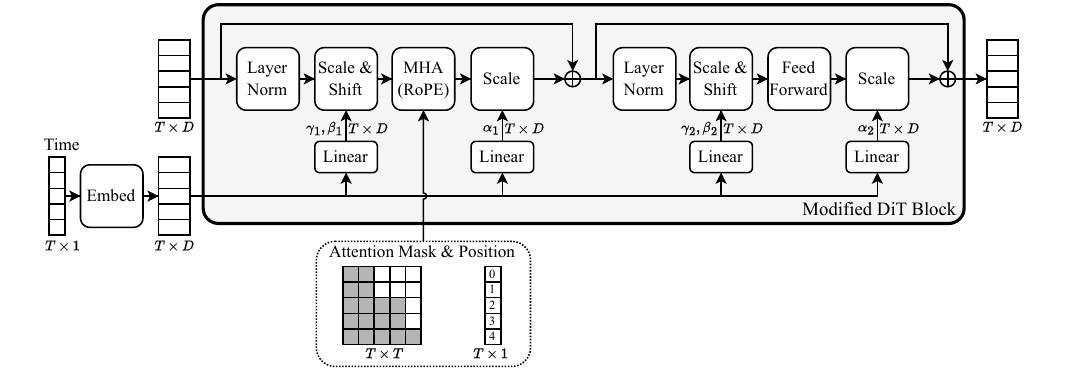}
    \caption{Modified DiT Block, with RoPE and individual timing for each token.}
    \label{fig:dit_block_modified}
\end{figure}

We have modified the DiT Block\footnote{DiT: \url{https://github.com/facebookresearch/DiT}} proposed in \cite{DiT} for 1D sequence processing, as illustrated in Figure~\ref{fig:dit_block_modified}. We replaced the original position embedding with RoPE, and added attention masks. The adaLN-Zero in the original DiT Blocks assumes a globally shared time. However, in our implementation, different times can be assigned to input tokens.

For the Mel spectrogram encoder, the Mel spectrogram is linearly projected to $\R^{N_{\text{frame}} \times D}$ before it is fed into the Transformer. To obtain the mean and log variance, the Transformer output is downsampled 4 times and linearly projected to $\R^{N_{\text{frame}} / 4 \times D_{\text{latent}}}$.

In the ARDiT model, the phoneme sequence is one-hot encoded and then linearly projected to $\R^{N_{\text{phone}} \times D}$. Similarly, the speech tokens are linearly projected to $\R^{N_{\text{latent}} \times D}$. These are then concatenated before being fed into the Transformer. To obtain the output velocity field, the Transformer output is segmented and linearly projected to blocks of shape $\R^{B \times D_{\text{latent}}}$. We assign a fixed time $t = -1$ to phoneme tokens, $t=0$ to clean tokens, and $t \in (0, 1]$ to noisy tokens.

\begin{figure}[ht]
    \centering
    \includegraphics[width=1.0\textwidth,trim=1.0cm 0cm 1.0cm .2cm,clip]{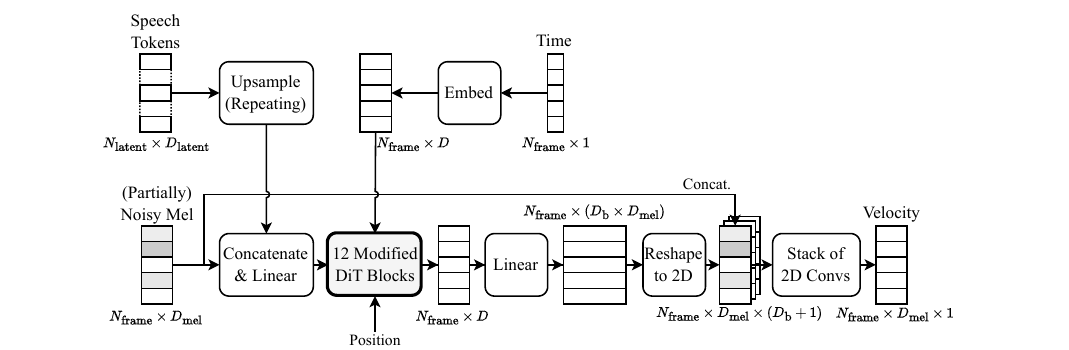}
    \caption{Mel spectrogram decoder architecture.}
    \label{fig:mel_decoder}
\end{figure}

We have specifically designed the Mel spectrogram decoder to enhance the generation of 2D Mel spectrograms. The structure of the decoder is depicted in Figure~\ref{fig:mel_decoder}. The output of the Transformer is projected and reshaped, then concatenated with the input noisy Mel spectrogram. This is further processed by a stack of 2D convolutions to obtain the velocity field. The stack of 2D convolutions consists of 6 Conv2D layers, with $D_b + 1 = 5$ input channels, $D_c = 128$ intermediate channels, and a single output channel. It features residual connections from the output of the 1st layer to the 3rd, and from the 3rd to the 5th. The kernel size for all layers is 3. All convolutions, except the last one, are followed by a leaky ReLU activation.

\section{Total Duration Estimation}

\label{appendix:duration}

The ARDiT TTS model depends on a given total duration of target speech during inference. There are many possible methods for modeling total duration. In our experiments, we used the following simple heuristic to estimate the total duration of the target speech.

First, obtain the average duration $\rho$ of non-silence phonemes. In our implementation, we force-aligned the reference speech and reference phoneme sequence using the ARDiT TTS model's attention matrix. Alternatively, $\rho$ can also be estimated using voice activity detection (VAD) on the reference speech. The estimated total duration is then $t' = \rho \cdot n'$, where $n'$ is the number of phonemes, excluding silences, in the target text. We also added additional time to $t'$ for punctuations in the target text, according to their average duration in spoken English.

We found that this rudimentary model performed reasonably well on the LibriTTS test set. A more principled method would involve training a conditional duration predictor \cite{VoiceBox}. However, this would require forced alignment to obtain the phoneme durations. We leave the exploration of improved total duration prediction for future work.

\section{Training Details} \label{sec:training_details}

The training of ARDiT TTS on LibriTTS involves several steps and can be completed within a week with 4 GPUs (NVIDIA RTX 4090). The steps are as follows:

\begin{enumerate}[label=\bfseries Step \arabic*:,leftmargin=*,labelindent=1em]
    \item Train the Mel spectrogram Autoencoder. This process takes approximately 3 days using 4 GPUs.
    \begin{itemize}
        \item The average batch size was 8.0 minutes per step.
        \item The model was trained for 590k training steps on 4 GPUs.
        \item The training speed was approximately 8.0k steps per hour.
    \end{itemize}
    \item Cache the latent code of all audios on the dataset for efficiency. This process takes approximately 1 hour using 1 GPU.
    \item Train the autoregressive diffusion transformer on TTS. This process takes approximately 2 days using 4 GPUs.
    \begin{itemize}
        \item The average batch size was 8.0 minutes per step.
        \item The training comprised 450k steps for ARDiT (B=1) and 640k steps for ARDiT (B=4).
        \item The training speed was approximately 11.3k steps per hour.
    \end{itemize}
    \item Cache noise and data pairs for efficiency. This process takes approximately 2 hours using 4 GPUs. For each utterance in the dataset, we cached one ODE trajectory for each block in the utterance (see Algorithm~\ref{alg:regression_loss}). The ODE sampling was done in parallel in all the blocks. We applied a 16 steps Euler sampler with fixed step size for sampling ODE trajectories.
    \item Distill ARDiT TTS models with DMD. This process takes approximately 2 days using 1 GPU.
    \begin{itemize}
        \item The average batch size was 1.2 minutes per step.
        \item ARDiT (DMD, B=1) was first trained with $\beta_{\text{reg}} = 2.0$ for 190k steps, then with $\beta_{\text{reg}} = 0.1$ for 10k steps.
        \item ARDiT (DMD, B=4) was first trained with $\beta_{\text{reg}} = 1.0$ for 200k steps, then with $\beta_{\text{reg}} = 0.2$ for 100k steps.
    \end{itemize}
\end{enumerate}

All models were optimized using the AdamW optimizer with $\beta=[0.9,0.95]$~\cite{AdamW}, a fixed learning rate of $0.0001$, and an Exponential Moving Average (EMA) with a decay rate of $0.9999$. The impact of $\beta_{\text{reg}}$ trajectories on performance in DMD training is left for future research.

\section{Listening Test with MUSHRA}\label{sec:mushra}

Each group of samples was initially evaluated by at least 25 participants. After applying quality filters and consistency checks, responses from 20 participants were considered for the final analysis. 

Each participant received a compensation of \$15 for their participation.

\begin{figure}[ht]
    \centering
    \fbox{\includegraphics[width=0.9\linewidth]{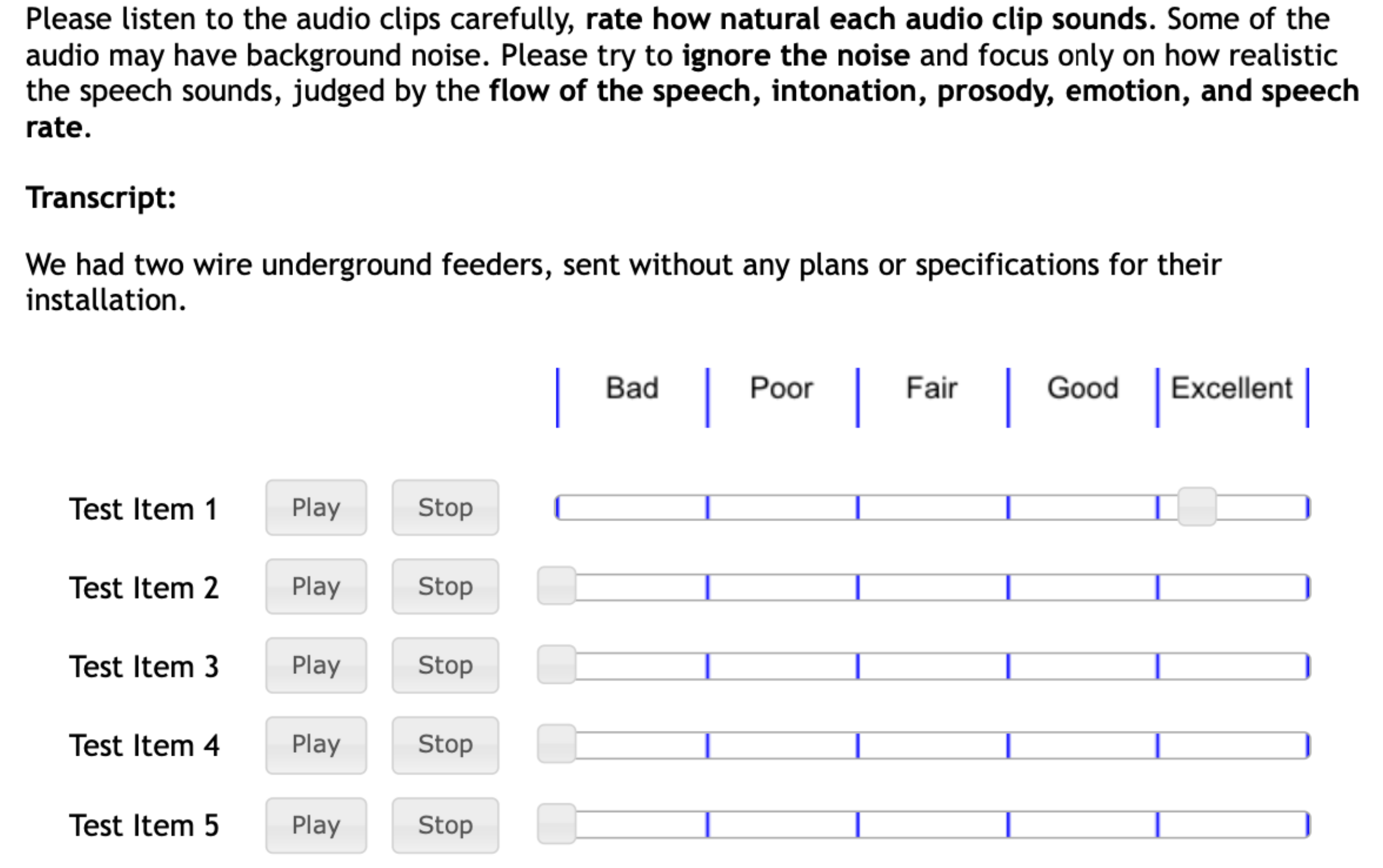}}
    \caption{Screenshot of speech naturalness test for zero-shot TTS.}
    \label{fig:mushra_nat}
\end{figure}

\begin{figure}[ht]
    \centering
    \fbox{\includegraphics[width=0.9\linewidth]{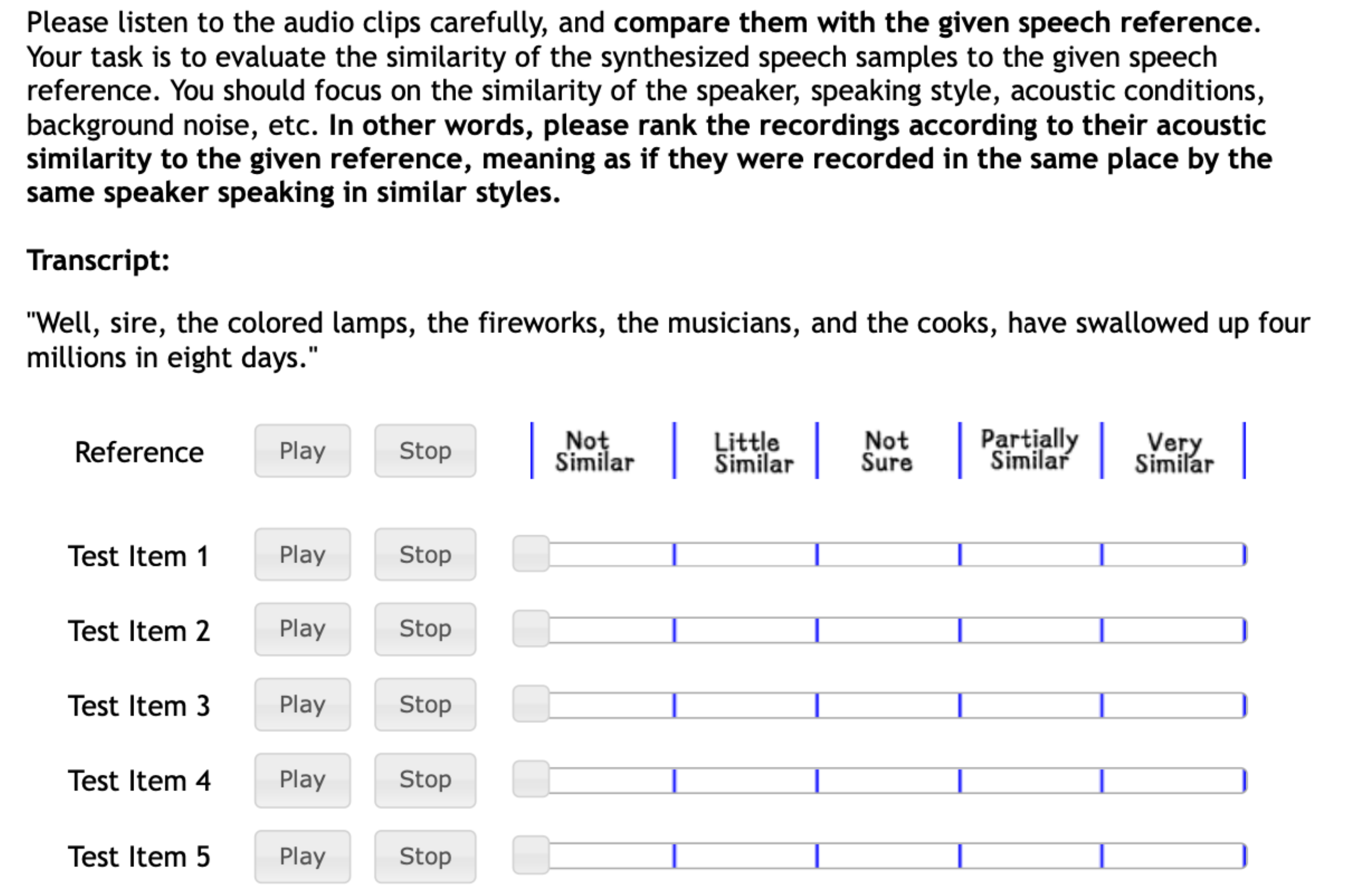}}
    \caption{Screenshot of speaker similarity test for zero-shot TTS.}
    \label{fig:mushra_sim}
\end{figure}

\begin{figure}[ht]
    \centering
    \fbox{\includegraphics[width=0.9\linewidth]{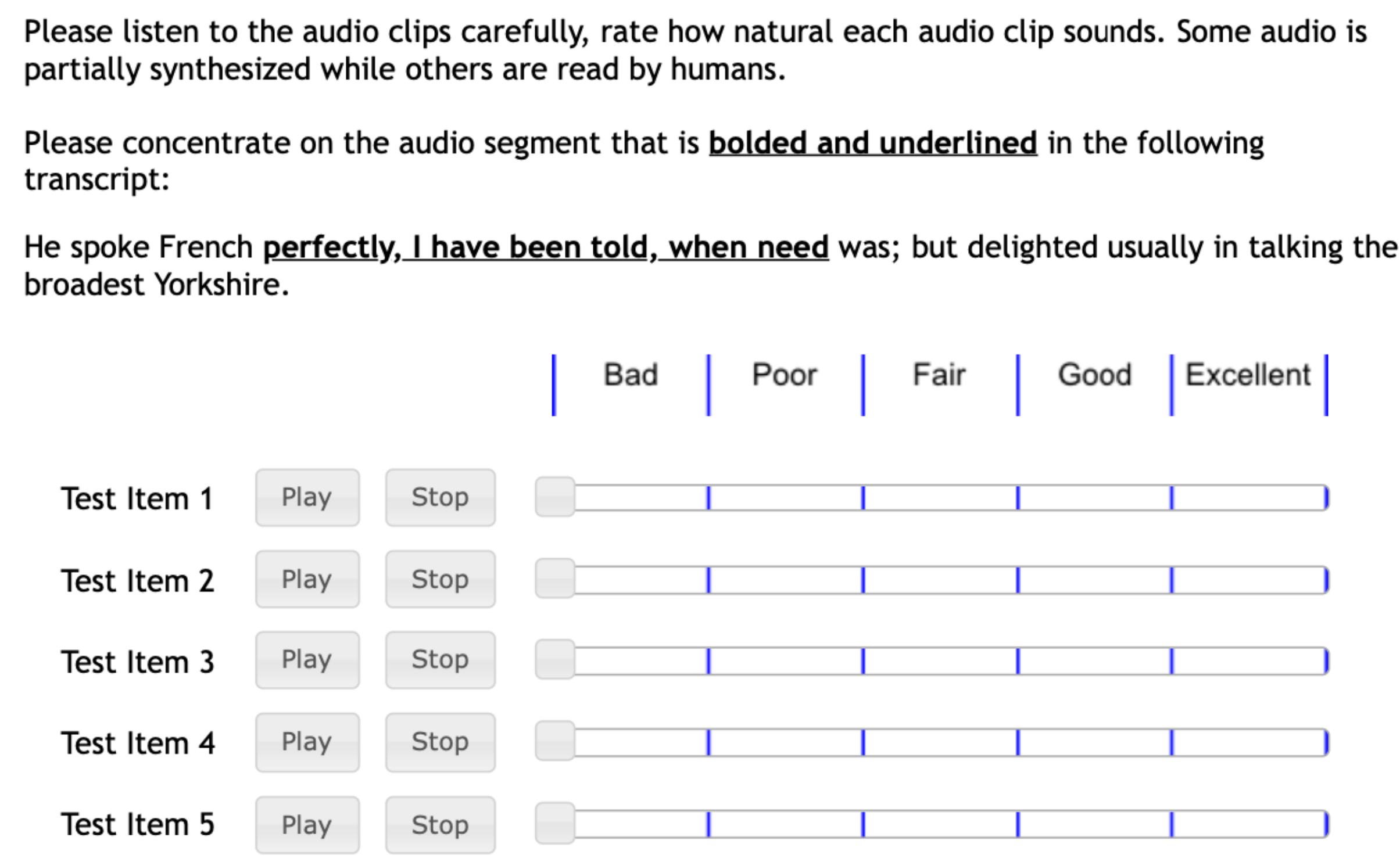}}
    \caption{Screenshot of speech naturalness test for speech editing.}
    \label{fig:mushra_edit_nat}
\end{figure}

For each participant in the listening test, we mandated that the survey be completed in a quiet environment using headphones. Detailed instructions with test examples are provided in Figure~\ref{fig:mushra_nat}, ~\ref{fig:mushra_sim} and ~\ref{fig:mushra_edit_nat}, which are the screenshots of the webpages shown to the listeners during the test~\cite{BeaqleJS}.

\section{Broader Impact}

Given that our model enables high-quality zero-shot speech synthesis and nearly perfect speech editing, it can benefit many related applications such as voice assistants, content creation, voiceovers, and more. However, it is important to note that this technology might present risks of misuse for deepfake audios or infringe on others' privacy due to high-quality voice cloning. From a technical perspective, it could be useful to incorporate watermarking or other technical safeguards to avoid this kind of misuse.

\clearpage
\end{document}